\documentclass[preprint,amsmath,amssymb,pra,aps]{revtex4-2}

\usepackage{graphicx}
\usepackage{bm}

\begin{document}



\title{Evolution of the velocity distribution of atoms\\ under the action of
  the bichromatic force}

\author{V. I. Romanenko}
 \altaffiliation{Institute of Physics of the National Academy of Sciences
   of Ukraine,\\  Nauky Avenue 46, Kyiv 03680, Ukraine}
 \email{vr@iop.kiev.ua}
 \author{L. P. Yatsenko}%
 \affiliation{Institute of Physics of the National Academy of Sciences
  of Ukraine,\\  Nauky Avenue 46, Kyiv 03680, Ukraine}%


\begin{abstract}
  We study numerically the evolution of the velocity distribution of
  atoms under the action of the bichromatic force. The comparison of
  the time dependencies of the distribution width and the average
  acceleration of atoms reveals the correlation of these
  quantities. We show that the estimation of the momentum diffusion
  coefficient on the basis of the analogy between the interaction of
  atoms with the counter-propagating bichromatic waves and the
  interaction of atoms with the counter-propagating sequences of the
  $\pi$-pulses roughly corresponds to the results of numerical
  calculations.  To separate the influence of the momentum diffusion
  on the evolution of atomic momentum distribution from the influence
  of the time-dependent Doppler shift, we study the motion of a
  ``heavy'' atom, for which the velocity change during the interaction
  of an atom with the field can be neglected.  Provided that the
  parameters of the atom-field interaction are optimal, we show that
  the momentum diffusion coefficient is proportional to the intensity
  of the laser radiation.  We used the Monte Carlo wave-function
  method for the numerical simulation of the atomic motion.
\end{abstract}

\maketitle

\tableofcontents

\section{Introduction}
\label{sec:Introduction}

The first theoretical study of the light pressure force exerted on
atoms in the field of counter-propagating bichromatic waves in
1988~\cite{Voitsekhovich-88} showed that this force can be used to
control the motion of atoms.  This control is based on the large light
pressure force exerted on atoms, which is much larger than the maximal
light pressure force $F_{sp}$ in the field of a traveling
monochromatic wave, given by the formula~\cite{Minogin,Metcalf-99}
\begin{equation}
  \label{eq:Fsp}
  F=\frac{1}{2}\hbar{}k\gamma.
\end{equation}
Here $\gamma$ is the rate of the spontaneous emission, $k$ is the wave
vector.  This fundamental limit was exceeded already in the first
observation of the bichromatic force~\cite{Voitsekhovich-89}.  Later,
the light pressure force much greater than $F_{sp}$was experimentally
confirmed~\cite{Soding}.  Analytical theory of the bichromatic force
was developed in the works~\cite{Yatsenko,Podlecki:18}, a review of
publications on this topic are given in~\cite{Metcalf}.  At the same
time, momentum diffusion of atoms in the field of counter-propagating
bichromatic waves has not been studied so far, in contrast to
well-known thorough studies of momentum diffusion of atoms in the
field of monochromatic waves~\cite{Minogin,Berg-Sorenson} and momentum
diffusion of atoms in the field of $\pi$-pulses~\cite{JETP}.

In this paper, we obtain statistical characteristics of the atomic
ensemble in the field of bichromatic force by numerical simulation of
the atomic state vector of each atom using the Monte Carlo wave
function method~\cite{Dalibard, Molmer}, followed by averaging over
all the atoms in the ensemble. We show that only at the beginning of
the interaction of the atoms with the field the time evolution of the
width $\Delta p$ of the momentum distribution of the atoms can be
described by the diffusion-like dependence $\Delta p = \sqrt{2Dt}$,
where $D$ is the momentum diffusion coefficient and $t$ is the
atom-field interaction time.  Over time, the width $\Delta p$ can both
increase and decrease depending on the change of the average velocity
of atoms.

To separate the influence of momentum diffusion on the evolution of
momentum distribution of atoms from the influence of the Doppler shift
of the atomic frequency, we used an assumption of a `` heavy '' atom,
when the change of velocity during the interaction of atom with the
field can be neglected. This allowed us to study the dependencies of
the momentum diffusion coefficient on the average velocity of atoms
and the momentum diffusion coefficient on the intensity of laser
radiation.

The paper is structured as follows. The next section presents the
equations that describe the time dependence of the field acting on the
atoms and the Hamiltonian of the atom-field interaction. The third
section describes the Monte Carlo method for the wave function. In the
fourth section, we present the scheme of the numerical
calculation. The obtained results are discussed in the fifth
section. In the sixth section, we formulate brief conclusions of the
work.

\section{Electric field and Hamiltonian}

Let's consider a two-level atom with the ground $|g\rangle $ and the
excited $|e\rangle $ states, which interacts with two
counter-propagating bichromatic waves
\begin{eqnarray}
  \bm{E}_{p}&=&\frac{1}{2}\bm{e}E_{0}\cos
              \left[
              \left(\omega+
              \frac{1}{2}\Omega
              \right)t+\frac{\varphi}{2}-kz \right]
+ \frac{1}{2}\bm{e}E_{0}\cos
              \left[
              \left(\omega-
              \frac{1}{2}\Omega
  \right)t-\frac{\varphi}{2}-kz \right]
    \label{eq:Ep}
  \end{eqnarray}
and
  \begin{eqnarray}
\bm{E}_{m}&=&\frac{1}{2}\bm{e}E_{0}\cos
              \left[
              \left(\omega+
              \frac{1}{2}\Omega
              \right)t-\frac{\varphi}{2}+kz \right]
+  \frac{1}{2}\bm{e}E_{0}\cos
              \left[
              \left(\omega-
              \frac{1}{2}\Omega
  \right)t+\frac{\varphi}{2}+kz \right],
                \label{eq:Em}
  \end{eqnarray}
  where $E_{0} $ is the peak to peak amplitude of the waves which is
  assumed to be the same for all waves, $\bm{e} $ is the unit
  polarization vector, $\omega\pm \frac{1}{2}\Omega$ are the carrier
  frequencies with the mean frequency $\omega$ and the difference
  $\Omega$.
               
  Here we neglected the difference of the wave vectors
  $\sim(\Omega/\omega)k$; taking it into account leads to a change in
  the phase difference which is substantial at a distance of the order
  $(\omega/\Omega)\lambda\sim(10^{6}\div10^{7})\lambda\sim1$~m, where
  $\lambda$ is the wavelength of electromagnetic radiation. In our
  model, used in other
  works~\cite{Voitsekhovich-88,Voitsekhovich-89,JETP,Soding,Yatsenko,Corder:15,Corder},
  we describe the phase difference between the waves by the terms
  $\pm\varphi/2$.

  The Hamiltonian of an atom in the field
    \begin{eqnarray}
      \label{eq:E}
     \bm{E}=\bm{E_{p}}+\bm{E_{m}}
    \end{eqnarray}
has the form
\begin{equation}
  \label{eq:Ham}
  H=\frac{\hat{\bm{p}}^{2}}{2M}+\hbar\omega_{0}|e\rangle\langle{e}|-
  \hat{\bm{d}}\cdot\bm{E}.
\end{equation}
Here $\hat{\bm{p}}$ is the momentum operator of the atom, $M$ is the
atomic mass, $\omega_{0}$ is the transition frequency (energy
difference of the states $e$ and $g$), $\hat{\bm{d}}$ is the dipole
momentum operator.

The field (\ref{eq:E}) can be written as counter-propagating
amplitude-modulated waves
\begin{equation}
  \label{eq:field}
  \bm{E}=\bm{e}E_{1}\cos(\omega t-kz)+\bm{e}E_{2}\cos(\omega t+kz),
\end{equation}
where
\begin{eqnarray}
  {E}_{1}& =& {E}_{0}\cos
  \left(
    \frac{1}{2}\Omega t + \frac{1}{2}\varphi
  \right);  \label{eq:field1}\\
{E}_{2}&=&{E}_{0}\cos
  \left(
    \frac{1}{2}\Omega t - \frac{1}{2}\varphi
  \right).
  \label{eq:field2}
\end{eqnarray}
Here $\frac{1}{2}\Omega$ is the modulation frequency (half the
difference of the frequencies of monochromatic waves $\Omega$ that
form bichromatic waves (\ref{eq:Ep}), (\ref{eq:Em})). The field
(\ref{eq:E}) can also be interpreted as a bichromatic standing wave
\begin{eqnarray}
  {E}& =& {E}_{0}\cos
         \left(
          kz+\frac{\varphi}{2}
          \right)\cos
          \left[
          \left(\omega-
          \frac{1}{2}\Omega   \right)t
          \right]
     +{E}_{0}\cos
          \left(
          kz-\frac{\varphi}{2}
          \right)\cos
          \left[
          \left(\omega+
          \frac{1}{2}\Omega   \right)t
          \right].
  \label{eq:field-st}
\end{eqnarray}

\section{Schr\"odinger equation and modeling of the state vector by
  Monte Carlo wave function method}
\label{sec:Schrodinger}

We determine the temporal evolution of the state vector
$\left|\psi\right\rangle$ from the Sch\"o{}dinger equation
\begin{equation}
  i\hbar\frac{d}{dt}\left| \psi\right\rangle=H\left| \psi\right\rangle-
i\hbar\frac{\gamma}{2}\left|e\right\rangle\left\langle{}e|\psi\right\rangle
\label{eq:Schrod}
\end{equation}
by the Monte Carlo wave function method~\cite{Dalibard,Molmer}.

The last term in Eq.~(\ref{eq:Schrod}) describes the spontaneous
emission by the atom in the excited state with the rate $\gamma$. To
reduce the number of equations needed to describe the evolution of the
state vector, we assume, as in~\cite{Molmer}, that the momentum of the
atom along the $Oz$ axis is changed after spontaneous emission by
$\pm\hbar{}k$ or does not change at all (the photon is emitted in the
orthogonal to $Oz$ axis direction).

 We seek for the state vector in the form
\begin{equation}
  \left| \psi\right\rangle =
  c_{g}(z,t)\left|g\right\rangle+c_{e}(z,t)e^{-i\omega_{0}t}\left|e\right\rangle.
\label{eq:psi}
\end{equation}
Substituting~\eqref{eq:field} and \eqref{eq:psi} in~\eqref{eq:Schrod},
we find the equations for $c_ {g}(z,t)$, $ c_ {e} (z, t)$ in the
rotating wave approximation~\cite{Shore}:
\begin{eqnarray}
  \label{eq:cg}
  i\frac{\partial}{\partial{}t}c_{g}&=&-\frac{\hbar}{2M}
                                        \frac{\partial^{2}}{\partial{}z^{2}}c_{g}+
                                           \frac{1}{2}\left(
                                           V_{1}e^{i\eta_{1}}+V_{2}e^{i\eta_{2}}
                                           \right)c_{e}, \\
  i\frac{\partial}{\partial{}t}c_{e}&=&-\frac{\hbar}{2M}
                                        \frac{\partial^{2}}{\partial{}z^{2}}c_{e}+
                                           \frac{1}{2}\left(
                                           V_{1}e^{-i\eta_{1}}+V_{2}e^{-i\eta_{2}}
                                             \right)c_{g}
                                        -i\frac{\gamma}{2}c_{e}.
  \label{eq:ce}
  \end{eqnarray}
  Here $\eta_{1}=(\omega-\omega_{0})t-kz$,
  $\eta_{2}=(\omega-\omega_{0})t+kz$ and $V_{1}$, $V_{2}$ are defined
  by expressions
\begin{eqnarray}
  \label{eq:V1}
  V_{1}&=&-\frac{1}{\hbar}\langle{}g|\hat{\bm{d}}\cdot{\bm{e}}|e\rangle{}E_{1},\\
  V_{2}&=&-\frac{1}{\hbar}\langle{}g|\hat{\bm{d}}\cdot{\bm{e}}|e\rangle{}E_{2}.
  \label{eq:V2}
\end{eqnarray}
To simplify the notations, in \eqref{eq:cg}, \eqref{eq:ce} and
hereafter we omit the arguments denoting the dependence of quantities
on time and coordinates.

We write $c_{g}$, $c_{e}$ in the form
\begin{eqnarray}
  \label{eq:cgb}
  c_{g}&=&\sum\limits_{-\infty}^{\infty}b_{g,n}\exp
           \left(
           {\frac{i}{2}\delta{t}+ik_{0}z-i\frac{\hbar{}k_{0}^{2}}{2M}t}
           \right)\langle{}z|n\rangle, \\
  c_{e}&=&\sum\limits_{-\infty}^{\infty}b_{e,n}\exp
           \left(
           {-\frac{i}{2}\delta{t}+ik_{0}z-i\frac{\hbar{}k_{0}^{2}}{2M}t}
           \right)\langle{}z|n\rangle,
  \label{eq:ceb}
\end{eqnarray}
where $\delta=\omega-\omega_{0}$ is the detuning of the average
frequence of the bichromatic wave from the transition frequency in the
atom, $\langle{}z|n\rangle=e^{inkz}$, $k_{0}=p_{0}/\hbar$, $ p_{0} $
is $z$ component of the initial momentum of the atom along the axis
$Oz$.  Time-dependent phases in Eqs.~\eqref{eq:cgb}, \eqref{eq:ceb} do
not influence the probabilities $b_{g,n}$, $ b_{e,n}$ to find the atom
in the states $|g,n\rangle = |g\rangle\otimes|n\rangle$,
$|e,n\rangle = |e\rangle\otimes|n\rangle$.

To obtain the equations for $b_{g,n}$, $b_{e,n}$, we substitute
equation~\eqref{eq:cgb}, \eqref{eq:ceb} in equations~\eqref{eq:cg},
\eqref{eq:ce}:
\begin{eqnarray}
  \frac{\partial}{\partial{}t}b_{g,n}&=&-i\left(n^{2}\delta_{rec}+
\frac{n\hbar{}kk_{0}}{M}+ \frac{1}{2}\delta
                                          \right)b_{g,n} 
-\frac{i}{2}\left(
                                          V_{1}b_{e,n+1}+V_{2}b_{e,n-1}
                                          \right) \label{eq:bg}\\
  \frac{\partial}{\partial{}t}b_{e,n}&=& -i\left(n^{2}\delta_{rec}+
\frac{n\hbar{}kk_{0}}{M} -\frac{1}{2}\delta
                                          \right)b_{e,n}
-\frac{i}{2}\left(
                                          V_{1}b_{g,n-1}+V_{2}b_{g,n+1}
                                          \right)-\frac{\gamma}{2}b_{e,n}.
  \label{eq:be}
\end{eqnarray}
Here $\delta_{rec}=\hbar{}k^{2}/(2M)$.

Hereinafter we assume that
$\langle{}g|\hat{\bm{d}}\cdot\bm{e}_{1}|e\rangle{}=
\langle{}g|\hat{\bm{d}}\cdot\bm{e}_{2}|e\rangle{}$, then
\begin{eqnarray}
  \label{eq:V1t}
  V_{1}&=&\Omega_{0}\cos
  \left(
    \frac{1}{2}\Omega t + \frac{1}{2}\varphi
  \right)\\
  V_{2}&=&\Omega_{0}\cos
  \left(
    \frac{1}{2}\Omega t - \frac{1}{2}\varphi
  \right),
  \label{eq:V2t}
\end{eqnarray}
where
$\Omega_{0}=-\langle{}g|\hat{\bm{d}}\cdot{\bm{e}}|e\rangle{}E_{0}/\hbar$.

We seek for the state vector~\eqref{eq:psi} by the Monte Carlo wave
function method~\cite{Dalibard,Molmer}. This method, when applied to
the amplitudes of the probabilities $ b_{g,n}$, $b_{e,n}$ to find an
atom in the ground or excited states with momentum
$\hbar{}k_{0}+n\hbar{}k$, looks like the following.
\begin{enumerate}
\item We assume that at time $t$ the amplitudes
  $b_{g,n}(t)$, $b_{e,n}(t) $ are normalized:
  \begin{equation}
    \label{eq:norm}
\sum_{n=-\infty}^{\infty}
\left(
|b_{g,n}(t)|^{2}+|b_{e,n}(t)|^{2}
\right) =1.
\end{equation}
Knowing $b_{g,n}(t)$, $b_{e,n}(t)$, we find the values of
$b_{g,n}^{(1)}(t+\Delta{}t)$, $b_{e,n}^{(1)}(t+\Delta{}t)$ from
Eqs.~\eqref{eq:bg}, \eqref{eq:be} for a small time interval
$\Delta{}t $. The presence of a dissipative term in~\eqref{eq:Schrod}
leads to a violation of the state vector normalization, therefore with
the change of time $t\to{}t+\Delta{}t$ the equality~\eqref{eq:norm} is
violated.  For a difference
  \begin{equation}
    \label{eq:normnew}
1-\sum_{n=-\infty}^{\infty}
\left(
|b_{g,n}^{(1)}(t+\Delta{}t)|^{2}+|b_{e,n}^{(1)}(t+\Delta{}t)|^{2}
\right)=\Delta{}P,
\end{equation}
we have
  \begin{equation}
    \label{eq:DP}
\Delta{}P=\gamma{}\Delta{}t\sum_{n=-\infty}^{\infty}
|b_{e,n}^{(1)}(t)|^{2}.
\end{equation}
The meaning of this equality is obvious: the factor
$\sum_{n=-\infty}^{\infty}|b_{e,n}^{(1)}(t)|^{2} $ is the population
of the excited state of the atom. Therefore, $\Delta{}P$ is the
probability of an atom to emit a photon during the time interval
$\Delta{}t$.
\item To find whether there was a photon emission during the time
  $\Delta{}t$, we generate a random variable $\epsilon$ which is
  uniformly distributed between zero and one and compare it with
  $\Delta{}P$. If $\epsilon>\Delta{}P$ (in most cases), no emission of
  photon has occurred. Then we form the probability amplitudes
  $b_{g,n}(t+\Delta{}t)$, $ b_{e,n}(t+\Delta{}t) $ at time
  $t+\Delta{t}$ by normalizing the values
  $b_{g,n}^{(1)}(t+\Delta{}t)$, $b_{e,n}^{(1)}(t+\Delta{}t) $ that
  where found in the first stage:
  \begin{eqnarray}
    \label{eq:bgtdt}
b_{g,n}(t+\Delta{}t)&=&\frac{b_{g,n}^{(1)}(t+\Delta{}t)}{\sqrt{1-\Delta{}P}},
    \\
b_{e,n}(t+\Delta{}t)&=&\frac{b_{e,n}^{(1)}(t+\Delta{}t)}{\sqrt{1-\Delta{}P}}.
        \label{eq:betdt}
  \end{eqnarray}
  If $\epsilon\leq\Delta{}P$, the atom emits a photon and goes to the
  ground state. In this case, after the emission of a photon
  $b_{g, n} (t+\Delta{}t)$ and $b_{e,n}(t+\Delta{}t)$ are defined as
  follows:
  \begin{eqnarray}
    \label{eq:bgtdts}
    b_{g,n}(t+\Delta{}t)&=&\frac{b_{e,n-\xi}^{(1)}(t+\Delta{}t)}{
                            \sum\limits_{m=-\infty}^{\infty}
                            |b_{e,m}^{(1)}(t+\Delta{}t)|^{2}}\\
    \\
    b_{e,n}(t+\Delta{}t)&=&0,
        \label{eq:betdts}
  \end{eqnarray}
  where $\xi$ takes one of the values $\xi = 0,\pm1$ with some
  probability.  Recall, that we simulate the real distribution of the
  projections of the photon momentum on the $Oz$ axis by a
  hypothetical distribution when $\xi = 0,\pm1$ (as if atoms emit
  photons either in the direction of the $ Oz $ axis or perpendicular
  to it), as was done in the modeling of Doppler cooling
  in~\cite{Molmer,Chretien}. In this case, we have a discrete
  distribution of atoms according to the projection of the spontaneous
  photon momentum on the axis $Oz$ with the step $\hbar{}k$.

  Instead of going back to step 1 and continuing the calculation
  further, we adjust the numbering of the amplitudes
  $b_{g,n}(t+\Delta{}t)$ in order to reduce the required size of the
  amplitude arrays in numerical calculations. A monotonic change of
  the average momentum of an atom over time (in the classical
  description of the motion of an atom, a force acts on it) increases
  the size of arrays of the probability amplitudes required for
  calculations. To reduce the size of the arrays, we calculate the
  average momentum acquired by the atom between the moments of
  spontaneous photon emission
\begin{equation}
  \label{eq:pat}
  \left\langle\Delta p\right\rangle=\hbar{}k\sum\limits_{n=-\infty}^{\infty}n
  \left|b_{g,n}\right|^{2},
\end{equation}
and find the integer number $N$ of photon momentum $\hbar{}k$ which it contains
\begin{equation}
  \label{eq:N}
  N=\left[ \frac{\left\langle\Delta p\right\rangle}{\hbar k}\right],
\end{equation}
where square brackets denote an integer part of a number.  Next, we
change the numbering of the amplitudes
\begin{equation}
  \label{eq:num}
b_{g,n}\rightarrow b_{g,n-N}.
\end{equation}
This is equivalent to changing the momentum of an atom to
$-N\hbar{}k $, so we also change the momentum
\begin{equation}
  \label{eq:mom}
p_{0}\rightarrow p_{0}+N\hbar k,
\end{equation}
so that the distribution of atoms by pulses does not change, and
return to step 1.
\end{enumerate}

For definiteness, we assume that the atom in the ground and excited
states is characterized by complete momentums $\hbar{}J_{g}$,
$\hbar{}J_{e}$ with $J_{e} = J_{g} + 1$. In this case, the two-level
scheme of the atom-light interaction between the states
$ |g,m_{g}=J_{g}\rangle$ and $|e,m_{e}=J_{e}\rangle$ (these are the
states that we denoted for simplicity by $|g\rangle$, $|e\rangle$) is
realized when the atom interacts with circularly polarized light. The
optimal description of the momentum diffusion rate due to spontaneous
emission by a discrete distribution of the projection of the photon
momentum on the axis $Oz$ occurs if $\xi$ acquires the values $ -1 $,
$ 0 $, $ + 1 $ with probabilities of $ 1/5 $, $ 3/5 $,
$ 1/5 $~\cite{Molmer,Chretien}. In essence, this means that the
specified distribution law for $\xi$ gives the same mean value of the
square of the projection of the photon momentum on the $Oz$ axis as
the real distribution of the projection of the photon momentum on the
$Oz$ axis.

In the calculations, we will also use the model of the heavy atom,
$M\to\infty$. In this case, the equations~\eqref{eq:bg}, \eqref{eq:be}
read
\begin{eqnarray}
  \frac{\partial}{\partial{}t}b_{g,n}&=&
                                         \left(
                                         -inkv-\frac{i}{2}\delta{}
                                         \right)b_{g,n}
-\frac{i}{2}\left(V_{1}b_{e,n+1}+V_{2}b_{e,n-1}\right) \label{eq:bg-inf},\\
  \frac{\partial}{\partial{}t}b_{e,n}&=&
                                         \left(
                                         -inkv+ \frac{i}{2}\delta{}
                                         \right)b_{e,n}
-\frac{i}{2}\left(V_{1}b_{g,n-1}+V_{2}b_{g,n+1}\right)-\frac{\gamma}{2}b_{e,n},
  \label{eq:be-inf}
\end{eqnarray}
where $v$ is the velocity of the atom which in the approximation
$ M\to\infty$ does not depend on time.

\section{Numerical calculation routine}
\label{sec:numerical}

We consider an ensemble of $N$ atoms in the field of the bichromatic
counter-propagating waves and assume that each atom begins to move
with a projection $v_{0}$ of the initial velocity on the axis
$Oz$. The evolution of the state vector of the atom is calculated by
the procedure described in section~\ref{sec:Schrodinger}. We repeat it
many times until we reach the final time of calculation $t_{f}$.

Knowing the final state vector of each of the $N$ atoms, we
determine the average momentum of the $m$-th atom by the formula
\begin{equation}
  \label{eq:p}
  \left\langle\! p^{(m)}\!\right\rangle=p_{0}^{(m)}+\hbar{}k\!\!\sum
  \limits_{n=-\infty}^{\infty}\!\!{} n
  \left(
    \left|
      b_{g,n}^{(m)}\right|^{2}+\left|b_{e,n}^{(m)}\right|^{2}
  \right),
\end{equation}
where the values of the probability amplitudes refer to the $m$-th
atom, and $p_{0}^{(m)}$ is the initial value of the momentum after the
last act of spontaneous radiation, which is modified after each
spontaneous photon radiation according to~\eqref{eq:mom}. The average
value of the square of the momentum of the $m$-th atom is calculated
by the formula
\begin{equation}
  \left
  \langle
  \left(
    p^{(m)}
  \right)^{2}\right\rangle=\sum\limits_{n=-\infty}^{\infty}
                              \left(p_{0}^{(m)}+n\hbar{}k\right)^{2}
 \left(
    \left|
      b_{g,n}^{(m)}\right|^{2}+\left|b_{e,n}^{(m)}\right|^{2}
  \right).
    \label{eq:pm}
\end{equation}
Now we can calculate the average value of $z$-component of the
momentum per an atom in the ensemble
\begin{equation}
  \label{eq:pav}
  p_{av}=\frac{1}{N}\sum\limits_{m=-\infty}^{\infty}\left\langle p^{(m)}\right\rangle
\end{equation}
and the standard deviation of the $z$-component of the momentum
\begin{equation}
  \label{eq:dp}
  \Delta{}p=\sqrt{\frac{\sum\limits_{m=-\infty}^{\infty}
      \left
        \langle
        \left(
          p^{(m)}
        \right)^{2}\right\rangle}{N}-p_{av}^{2}}
\end{equation}
per an atom from its average value. Eqs.~\eqref{eq:pav},
\eqref{eq:dp} allow us to calculate the average force $F$ acting on
the atom and the momentum diffusion coefficient $D$:
\begin{eqnarray}
  \label{eq:Ft}
  F&=&\frac{p_{av}}{t_{f}},\\
  D&=&\frac{\Delta{}p^{2}}{2t_{f}}
       \label{eq:Dt}
\end{eqnarray}
In the following calculations, we will compare the force of light
pressure   on the atom in the field of counter-propagating bichromatic
waves with the maximal force $F_{sp}$ of pressure on the atom in the field of
the monochromatic traveling wave given by Eq.~\eqref{eq:Fsp}. 
The momentum diffusion coefficient we will  compare with the maximal
  momentum diffusion coefficient $D_{r}$ in the
field of a traveling monochromatic wave along the direction of its
propagation~\cite{Minogin}:
\begin{equation}
  \label{eq:Da}
  D_{r}=\frac{1}{4}\hbar^{2}{}k^{2}\gamma(1+\alpha),
\end{equation}
where $\alpha=\langle\cos^{2}\theta\rangle $ is the mean value of the
square of the cosine of the angle between the direction of photon
radiation and the direction of wave propagation. For the model that we
adopted here $\alpha = 2/5$.

\section{Results of numerical simulations}
\label{sec:results}

We perform numerical simulation of temporal evolution of the average
velocity and the variance of the velocity of atoms in the field of
counter-propagating bichromatic waves for sodium and cesium atoms. In
addition, we consider the force of light pressure acting on atoms and
the variance of the velocity of atoms for the limit case of very heavy
atoms, when the change of the velocity of an atom (but not the
momentum of the atom) during its interaction with the field can be
neglected.

Calculations were carried out for atoms $^{23}\mathrm{Na}$ and
$^{133}\mathrm{Cs}$, in which a cyclic interaction with the field can
be created~\cite{Metcalf-99}. The wavelength of the transition
$3{}^{2}S_{1/2}-3{}^{2}P_{3/2}$ in the sodium atom is
$\lambda = 589.16$~nm, the rate of spontaneous emission is
$\gamma = 2\pi\times9.795$~MHz, the Doppler cooling limit is
$T_{D} = 235.03$~ $\mu$K~\cite{Steck:Na}. In a cesium atom, the
wavelength of the transition $6{}^{2}S_{1/2}-6{}^{2}P_{3/2}$ is
$\lambda = 852.35$~nm, the rate of spontaneous emission is
$\gamma = 2\pi\times5.23$~MHz, the Doppler limit of atomic cooling is
$T_{D} = 125.61$~$\mu$K~\cite{Steck:Cs}. All calculations were
performed for the case $\omega = \omega_{0}$, i.e. $\delta = 0$.

\subsection{Sodium and cesium atoms in the field of
  counter-propagating bichromatic waves}
\label{sec:quant-na}
\begin{figure}
  \includegraphics[width=86mm]{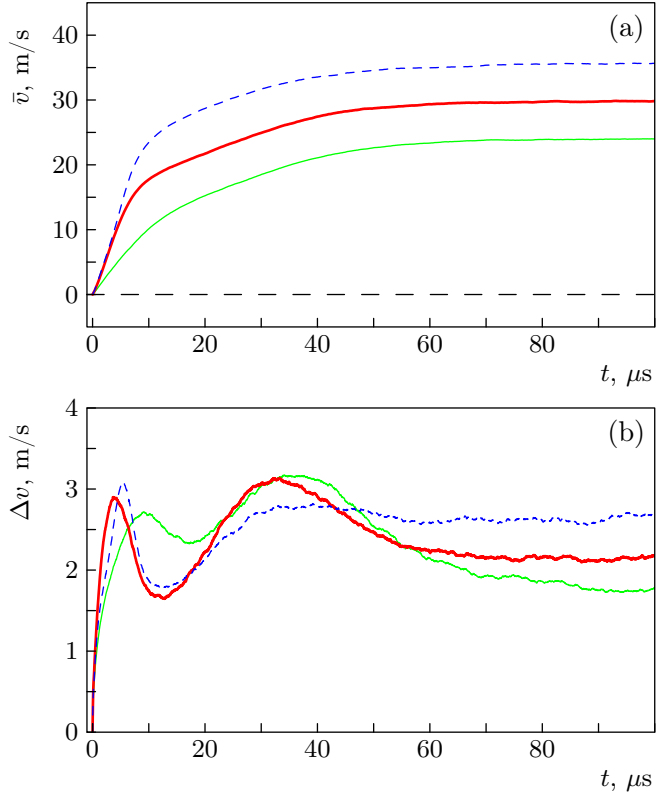}
  \caption{Time dependence of the average velocity
    $\bar{v}$ (a) and the square root of the velocity variance
    $\Delta{}v$ (b) for 1000 sodium atoms in the field of the
    counter-propagatiing bichromatic waves. Parameters are:
    $\Omega_{0}=2\pi\times122$~MHz; $\Omega = 2\pi\times80$~MHz (thin
    curve), $\Omega{=}2 \pi\times100$~MHz (thick curve),
    $\Omega = 2\pi\times120 $~ MHz (dashed curve). The initial
    velocity of the atoms is $v_{0}=0$. The phase difference of the
    counter-propagating waves is $\varphi=\pi/4$}
\label{fig-R122-v}
\end{figure}

Fig.~\ref{fig-R122-v} shows time evolution of the mean velocity
$\bar{v}$ and the standard deviation of the velocity $\Delta{}v$ from
its mean value (square root of the velocity variance) for sodium atoms
in the field of counter-propagating bichromatic waves. The plots are
obtained both for the optimal ratio of the Rabi frequency $\Omega_{0}$
to the difference of the frequencies of monochromatic waves $\Omega$
that form bichromatic waves at $\varphi = \pi/4 $
($\Omega_{0}=\sqrt{3/2} \, \Omega$~\cite{Yatsenko}) and for a small
but noticeable ($\approx20\%$) deviation from the optimum.
First of all, it should be noted that with increasing time the
velocity of the atom approaches $v=\Omega/(2k)$, when, according to
the quasi-classical theory~\cite{Voitsekhovich-88,Soding}, the force
of light pressure exerted on the atom is zero. The standard deviation
$\Delta{}v$ of velocity from the mean value of $\bar{v}$ increases
monotonically with time only at the beginning of the interaction of
atoms with the field, then this dependence becomes nonmonotonic. This
suggests that influence of the Doppler effect, which becomes more
noticeable with increasing velocity, on the distribution of atoms in
the momentum space is significant. Also, the Doppler effect leads to a
change in the almost linear dependence of velocity on time at the
beginning of the interaction with the field (which indicates an almost
constant value of the light pressure force exerted on the atom) to
nonlinear. In the next section, to exclude the influence of the
time-dependent Doppler shift on the light pressure force and the
momentum dispersion, we consider the interaction of atoms of very
large mass with counter-propagating bichromatic waves.

\begin{figure}[th]
    \includegraphics[width=86mm]{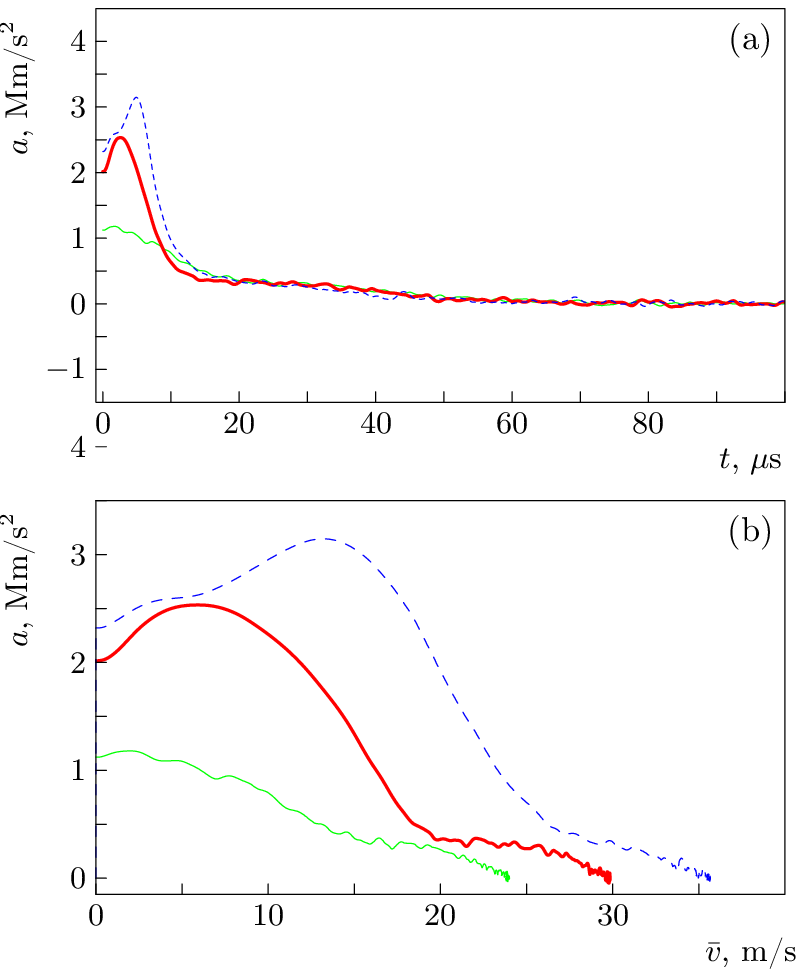}
  \caption{Time dependence of the average acceleration
    $a = d\bar{v}/dt$ (a) and the dependence of the average
    acceleration $a$ on the average velocity $\bar{v}$ of atom (b) for
    1000 sodium atoms in the field of counter-propagating bichromatic
    waves. Parameters are: $\Omega_{0}=2\pi\times122$~MHz;
    $\Omega=2\pi\times80$~MHz (thin curve),
    $\Omega{=}2\pi\times100$~MHz (thick curve),
    $\Omega=2\pi\times120$~MHz (dashed curve). The phase difference of
    the counter-propagating waves is $\varphi=\pi/4$ }
\label{fig-R122-a-va}
\end{figure}

Fig.~\ref{fig-R122-a-va},~\emph{a} shows the time dependence
of the acceleration $a{=}d\bar{v}/dt $, calculated by differentiation
of the time dependence of the average velocity, which is depicted in
Fig.~\ref{fig-R122-v}, \emph{a}, after its pre-smoothing with the
program \textsc{Gnuplot} (option \texttt{acsplines} with smoothing
parameter 1). Such smoothing is necessary because the derivative of
the unsmoothed dependence of $\bar{v}$ on time has peculiarities due
to quantum jumps in the atoms. A mechanical analog of this averaging
procedure is the movement of the piston in a cylinder filled with some
gas under the action of molecules. Note that the smoothed dependence
of $\bar{v}$ on time is visually indistinguishable from the
non-smoothed dependence of $\bar{v}$ which is shown in
Fig.~\ref{fig-R122-v}, \emph{a}.  The maximum average acceleration of
atoms in Fig.~\ref{fig-R122-a-va} is more than three times higher than
the maximum acceleration of the atom in the field of a traveling light
wave, which for the sodium atom is
$a_{sp}=F_{sp}/m=\hbar{k}\gamma/(2m) = 0.925\cdot10^{6}$~m/s$^2$,
which roughly corresponds to the results of \cite{Voitsekhovich-88}
obtained for the model of the ``heavy'' atom.

Comparing the time dependence of the average velocity $\bar{v}$ of
atoms against the smoothed time dependence of the acceleration, one
can obtain the dependence of the average acceleration $a$ of atoms on
the average velocity (see Fig.~\ref{fig-R122-a-va}, \emph {b}). The
fluctuation-like dependence of acceleration at high atomic velocities,
when the value of $k\bar{v}$ is close to $\Omega/2 $, is due to the
relatively small number of atoms in our calculations.

\begin{figure}[b]
  \includegraphics[width=86mm]{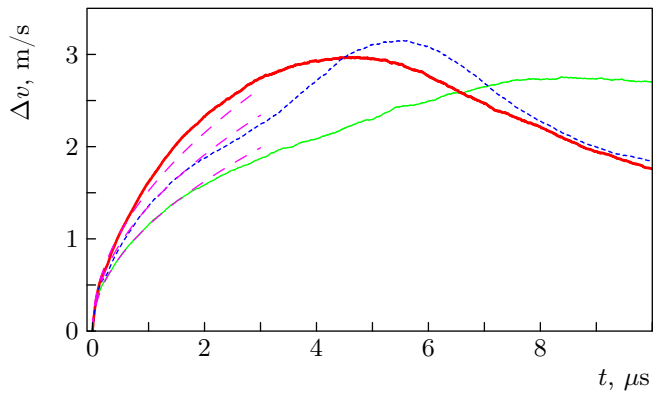}
  \caption{Time dependence of the standard deviation
    $\Delta{}v$ of the velocity of atoms on its average value
    $\bar{v}$ for 1000 sodium atoms in the field of
    counter-propagating bichromatic waves. Parameters are:
    $\Omega_{0}{=}2\pi\times122$~MHz; $\Omega=2\pi\times80$~MHz (thin
    curve), $\Omega/2\pi{=}100$~MHz (thick curve),
    $\Omega=2\pi\times120$~MHz (short dashed curve). The initial
    velocity of the atoms is $v_{0}=0$. The long dashed curve shows
    the dependencies $\Delta{}v=\sqrt{2D_{v}t}$, close to the curves
    with $\Omega{=}2\pi\times80$~MHz
    ($D_{v}=0.6584\cdot10 ^ {6}$~m$^{2}$/s$^{3}$),
    $\Omega=2\pi\times100$~MHz
    ($D_{v}=1.1562\cdot10^{6}$~m$^{2}$/s$^{3}$),
    $\Omega=2\pi\times120$~MHz
    ($D_{v}=0.9135\cdot10^{6}$~m$^{2}$/s$^{3}$).  Phase difference of
    the counter-propagating waves is $\varphi=\pi/4$ }
\label{fig-R122-dv}
\end{figure}

In Fig.~\ref{fig-R122-dv} we show the initial part of the dependence
of $\Delta{}v$ on $\bar{v}$ in Fig.~\ref{fig-R122-v}.
At the beginning of the interaction of atoms with the field, the
calculated dependencies of $\Delta{}v$ on $t$ are well approximated by
the curves $\Delta{}v=\sqrt{2D_{v}t} $. This indicates the diffusion
nature of the spreading of the distribution of atoms by momentum at
the beginning of their interaction with the field. For a larger time,
the dependence of $\Delta{}v $ on time is no longer described by the
diffusion law that can be explained by the dependence of the diffusion
coefficient $D_{v}$ on the velocity of the atom.

The diffusion coefficient $D_{v}$ in the velocity space is related to
the momentum diffusion coefficient $D$ by the relation:
\begin{equation}
  \label{eq:DDv}
  D_{v}=D/m^{2}.
\end{equation}
Let us compare the diffusion coefficient in the velocity space in the
field of counter-propagating bichromatic waves with that in the field
of a traveling monochromatic wave of high intensity (when there is a
saturation of absorption)
\begin{equation}
  \label{eq:DDvr}
  D_{vr}=D_{r}/m^{2},
\end{equation}
where $D_{r}$ is given by the expression~(\ref{eq:Da}). Calculations
for sodium atoms with $\alpha = 0.4$ give $ D_{r}=2.78\cdot10^{- 47}$
kg$^{2}$m$^{2}$/s$^{3}$, $D_{vr}=19.07\cdot10^{3}$m$^{2}$/s$^{3}$. For
the curves shown in Fig.~\ref{fig-R122-dv} we have the ratio
$D_{v}/D_{vr}$: 34.5 ($\Omega/2\pi=80$~MHz), 60.6
($\Omega/2\pi=100$~MHz), 47.9 ($\Omega/2\pi = 120$~MHz), i.e. the
coefficient of momentum diffusion of sodium atoms in the field of
counter-propagating bichromatic waves exceeds the coefficient of
momentum diffusion in the field of traveling monochromatic wave by
1--2 orders.

Now we compare the calculated momentum diffusion coefficients with a
rough estimate of the diffusion coefficient, which can be obtained by
a close analogy between the interaction of atoms with sequences of
counter-propagating $\pi$-pulses and the bichromatic
field~\cite{Voitsekhovich-88, Voitsekhovich-89, JETP, Soding,
  Yatsenko}.  In paper~\cite{JETP}, we found the coefficient of
diffusion of atoms in the field of counter-propagating sequences of
$\pi$-pulses with the repetition period $ T $ for the model of a
``heavy'' atom, the maximum value of which (for $\tau=\frac{1}{2}T$,
$\tau$ is the time shift between opposing pulses at the location of
the atom) in the most interesting case $\gamma{}T\ll{}1$ reaches
  \begin{equation}
    \label{eq:Dpi}
    D_{\pi\max}=\frac{4\hbar^{2}k^{2}}{\gamma{}T^{2}}.
  \end{equation}
  Maximal value of the light pressure force in the field of the
  counter-propagating bichromatic waves
  \begin{equation}
    \label{eq:bi}
    F_{bi}=\frac{\hbar{}k}{\pi}\Omega 
  \end{equation}
  may achieve the maximal light pressure force in the field of the
  sequences of the counter-propagating $\pi$-pulses with a repetition
  period $T=2\pi/\Omega$~\cite{Soding,Yatsenko}.  This formula is
  valid provided that the optimal parameters of interaction of atoms
  with the field are chosen~\cite{Yatsenko}, in particular, for
  $\Omega_{0}=\sqrt{\frac{3}{2}}\Omega$ and $\varphi=\frac{1}{4}\pi$,
  if the atom occupies only one of the ``dressed'' states. In the
  general case, the real light pressure force exerted on atoms, taking
  into account the distribution of atoms by states, can be calculated
  only numerically~\cite{Yatsenko}.

  For a rough estimation of the momentum diffusion coefficient in the
  field of the counter-propagating bichromatic waves, we use the
  analogy between the interaction of atoms with sequences of
  counter-propagating $\pi$-pulses and the interaction of atoms with
  the counter-propagating bichromatic waves. Substituting
  $T=2\pi/\Omega$ in~\eqref{eq:Dpi}, we obtain a rough estimation of
  the momentum diffusion coefficient of atoms in the field of the
  counter-propagating bichromatic waves:
    \begin{equation}
    \label{eq:Dbi}
    D_{bi}=\frac{\hbar^{2}k^{2}\Omega^{2}}{\gamma{}\pi^{2}}.
  \end{equation}
  For sodium atoms Eq.~\eqref{eq:Dbi} gives
  $D_{bi}=8.04\cdot10^{-46}$~kg$^{2}$m$^{2}$/s$^{3}$ for
  $\Omega=2\pi\times100$~MHz. Estimation of the momentum diffusion
  coefficient from Fig.~\ref{fig-R122-dv} for
  $\Omega=2\pi\times100$~MHz gives
  $D=1.7\cdot10^{-45}$~kg$^{2}$m$^{2}$/s$^{3}$, that is twice
  $D_{bi}$. The relatively small difference between $D$ and $D_{bi}$
  indicates the expediency of using   formula ~(\ref{eq:Dbi}) for  
  rough estimation of the momentum diffusion coefficient in the field
  of counter-propagating bichromatic waves provided that the optimal
  parameters of the atom-field interaction are chosen.

\begin{figure}[th]
  \includegraphics[width=86mm]{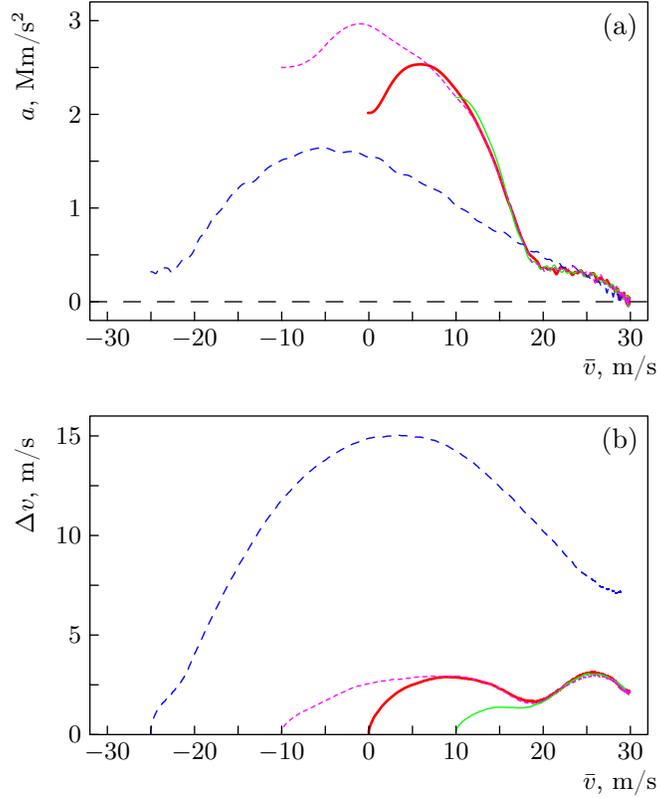}
  \caption{Dependence of the average acceleration
    $a=d\,\bar{v}/dt $ on the average velocity $\bar{v}$ of atoms,
    calculated after smoothing $\bar{v}$ by the program
    \textsc{Gnuplot} (a) and the dependence the standard deviation of
    the velocity $\Delta{}v$ from its mean value $\bar{v}$ on the
    average velocity $\bar{v}$ of atoms (b) for 1000 sodium atoms in
    the field of counter-propagating bichromatic waves for different
    values $ v_ {0} $ of the initial velocity of atoms. Parameters
    are: $\Omega_{0}=2\pi\times122$~MHz; $\Omega=2\pi\times100$~MHz;
    thick solid curve corresponds to $v_{0}=0$, thin solid curve
    corresponds to $v_{0}=10$~m/s, short dashed curve corresponds to
    $v_{0}=-10$~m/s, long dashed curve corresponds to
    $v_{0}=-25$~m/s. Phase difference of counter-propagating waves is
    $\varphi=\pi/4$}
\label{fig-R122-v-a}
\end{figure}

  To estimate the influence of the initial velocity $v_{0}$ on the
  mean acceleration $a$ of the atom's motion and the standard
  deviation of the velocity from its mean value, it is convenient for
  each of the pairs of dependencies $a(t)$, $\bar{v}(t)$ and
  $\Delta{v}(t)$, $\bar{v}(t)$, calculated for the motion of the
  ensemble of atoms in the field of counter-propagating bichromatic
  waves, to construct curves whose abscissas are equal to $\bar{v}$,
  and the ordinates are $a$ (Fig.~\ref{fig-R122-v-a}, \emph{a}), and
  $\Delta{}v$ (Fig.~\ref{fig-R122-v-a}, \emph{b}). To simplify the
  terminology, we will talk about these curves as dependences of $a$
  and $\Delta{}v$ on $\bar{v}$.
The physical basis of these dependencies is different conditions for
the interaction of atoms with the field due to the Doppler effect and
different initial conditions. Calculation of the average acceleration
of atoms is performed by differentiating the dependence of the average
velocity of atoms on time after its pre-smoothing using the program
\textsc{Gnuplot} (option \texttt{acsplines} with smoothing parameter
1). We also calculated the momentum diffusion coefficient for
$v_{0}=-25$~m/s, using the dependence $\Delta{}v(t)$ not given here
and obtained $D_{v}=0.34\cdot10^{6}$~m$^{2}$/s$^{3}$, which is half as
much as that at $v_{0}=0$, given in the caption to
Fig.~\ref{fig-R122-dv}.

Shown in Fig. \ref{fig-R122-v-a}, \emph{a} dependencies of $a$ on
$\bar{v}$ for $v_{0}=0,\pm10$~m/s in common areas of definition differ
a little, that is natural to expect, paying attention to the
dependencies $\Delta{}v(\bar{v})$ in
Fig.~\ref{fig-R122-v-a},~\emph{b}. The variation $k\Delta{}v$ of the
Doppler shifts of the frequency of atoms calculated using these
dependencies is much smaller than the average value of the Doppler
shift $k\bar{v}$, so the atoms in the ensemble have approximately the
same velocities. As far as the atoms with the initial velocity of
$v_{0}=-25$~m/s are concerned, the situation is completely
different. In this case the corresponding curve in
Fig.~\ref{fig-R122-v-a},~\emph{a} passes much lower than the curves
corresponding to $v_{0}=0,\pm10$~m/s. The reason is obvious~--- in the
ensemble of atoms there is a significant variation of $k\Delta{}v$
Doppler frequency shifts relative to the average Doppler shift
$k\bar{v}$ (see the corresponding curve in
Fig.~\ref{fig-R122-v-a},~\emph {b}), and this dramatically affects the
interaction of atoms with the field. In this case, it is manifested by
a marked decrease in the average acceleration of atoms.

Note that the oscillations on all three curves with
$v_{0}=0,\pm10$~m/s in Fig.~\ref{fig-R122-v-a}, \emph{a} for
$\bar{v}>20$~m/s, when the average acceleration of atoms is close to
zero, do not match. This suggests that the oscillations are due to
fluctuations when averaged over a relatively small number of atoms.

At the beginning of the interaction of atoms with the field, the
standard deviation of the velocity from the mean value changes
approximately according to the law $\Delta{}v\propto\sqrt{2D_{v}t}$
(see Fig.~\ref{fig-R122-dv}). Over time, the dependence of $\Delta{}v$
on time (as well as on the average velocity $\bar{v}$ of atoms, as far
as $\bar{v}$ grows monotonically with time) ceases to be monotonic:
are observed as areas of growth, and the decline of $\Delta{}v$ (see
Fig.~\ref{fig-R122-v-a}, \emph{b}).

To understand the nonmonotonic change of $\Delta{v}$ with a change of
$\bar{v}$, we must take into account two factors that determine it. On
the one hand, it is a process of momentum diffusion, which leads to a
diffusion-like change of $\Delta{}v$ with time, as can be seen in
Fig.~\ref{fig-R122-dv}. On the other hand, the dependence of the
average acceleration $a$ on the average velocity $\bar{v}$ of an atom
can both increase and decrease the value of $\Delta{}v$. Indeed, let
us consider the case $da/d\bar{v}<0$. It is plausible to assume that
the acceleration $a_{1}$ of one atom or group of atoms with velocity
$v_{1}$ also decreases with increasing velocity,
$da_{1}/d{v_{1}}<0$. The accuracy of this statement becomes greater if
the standard deviation $\Delta{}v$ of the velocity of atoms in the
ensemble from its mean value becomes smaller. In this case, the
acceleration of atoms with a lower velocity is greater than the
acceleration of atoms with a higher velocity. As a result, the
velocity distribution of atoms narrows to a limit that is established
as a result of the dynamic equilibrium of the distribution narrowing
process by reducing the acceleration of atoms with increasing velocity
and the momentum diffusion process that makes broader this
distribution. A well-known analog of this phenomenon is the Doppler
cooling of atoms~\cite{Metcalf}, with the difference that instead of
grouping atoms at zero velocity in the field of a standing
monochromatic wave, we have the grouping of atoms near the average
velocity of the ensemble of atoms $\bar{v}$ in the field of
counter-propagating bichromatic waves in the case of
$da/d\bar{v}<0$. Similar considerations show that in the case of
$da/d\bar{v}>0$ there should be an expansion of the distribution of
atoms by momentum. Besides that, in this case, the dependence of the
acceleration of atoms on their velocity changes $\Delta{}v$ in the
same direction as the momentum diffusion. The described phenomena are
clearly seen in Fig.~\ref{fig-R122-v-a}.

Now, understanding the physical reason for the nonmonotonic dependence
of $\Delta{}v$ on $\bar{v}$, we will dwell on some details of the
dependencies shown in Fig.~\ref{fig-R122-v-a}. First, we note the
shift of the maximum of $\Delta{}v(\bar{v})$ relative to $a(\bar{v})$
towards the larger $\bar{v}$, which is well visible for the curve
corresponding to initial velocity $v_{0}=-25$~m/s. The reason for this
shift is obvious. The derivative $da/d\bar{v}$ near the maximum of the
dependence $a(\bar{v})$ is too small to compensate the increase in
$\Delta{}v$ due to the momentum diffusion. The small value of the
derivative $da/d\bar{v}$ in the range from 20~m/s to 25~m/s,
insufficient to compensate the momentum diffusion, is also the reason
for the growth of $\Delta{}v $ for curves with $v_{0}=0$~m/s,
$v_{0}=5$~m/s and $v_{0}=10$~m/s in Fig.~\ref{fig-R122-v-a},
\emph{b}. For the curve with $v_{0}=-25$~m/s, the derivative
$da/d\bar{v}$ in the interval from 5~m/s to 30~m/s is large enough to
monotonically reduce the standard deviation of the velocities of atoms
from the mean value, and no local maxima to the right of $v_{0}=5$~m/s
are observed.

\begin{figure}[th]
  \includegraphics[width=86mm]{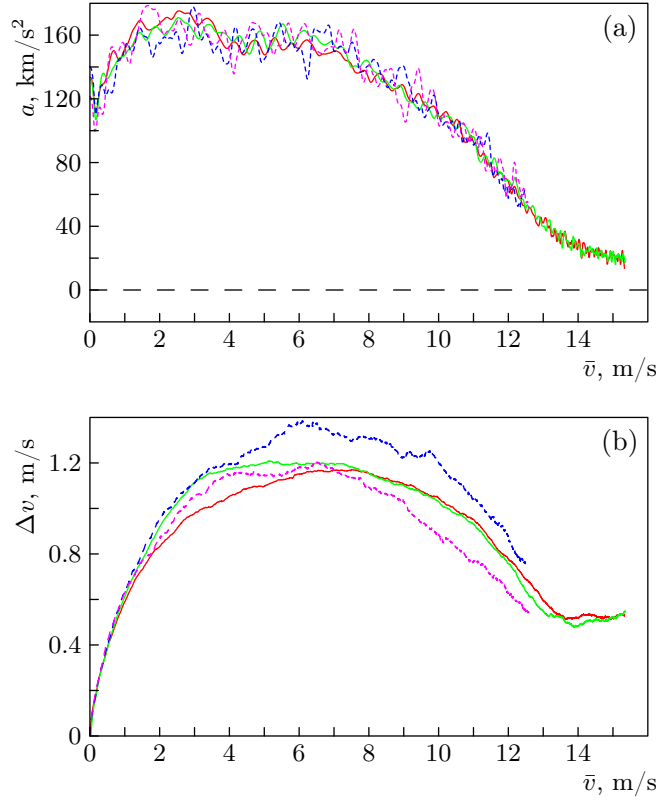}
  \caption{ Dependence of the average acceleration
    $a=d\,\bar{v}/dt$ on the average velocity of $\bar{v}$ cesium
    atoms, calculated after smoothing $\bar{v}(t)$ by the program
    \textsc{Gnuplot} (a) and dependence of the standard deviation of
    the velocity $\Delta{}v $ on its average value $\bar{v}$ (b) for
    500 (solid curve) and 100 (dashed curve) cesium atoms in the field
    of counter-propagating bichromatic waves for different
    implementations of sequences of random numbers. Parameters are:
    $\Omega_{0}=2\pi\times63.44$~MHz;
    $\Omega=2\pi\times{}51.8$~MHz. The initial velocity of atoms
    $v_{0}=0$, the phase difference of the opposing waves
    $\varphi=\ pi/4$. The calculations were performed for the atomic
    motion time of 100 $\mu$s in the case of 100 atoms and 200 $\mu$s
    in the case of 500 atoms}
  \label{fig-Cs-v-a-dv}
\end{figure}

Similar behavior of $\Delta{}v$ for $^{23}$Na atoms occurs also for
atoms $^{133}$Cs.  Fig.~\ref{fig-Cs-v-a-dv} shows the dependence of
the average acceleration $a=d\,\bar{v}/dt$ on the average velocity of
$\bar{v}$ atoms, calculated after smoothing $\bar{v}(t)$ by the
program \textsc {Gnuplot} with smoothing parameter 1 (\emph{a}) and
the dependence of the standard deviation of the velocity $\Delta{}v$
on its mean value $\bar{v}$ (\emph {b}).
The calculations were performed for the same values of
$\Omega_{0}/\gamma$, $\Omega/\gamma$, and $\varphi$, as well as for
sodium atoms in Fig.~\ref{fig-R122-v-a}. Again, as in
Fig.~\ref{fig-R122-v-a}, we see a decrease in $\Delta{}v $ with
decreasing of the average acceleration of atoms (see the range of
$\bar{v}>8$~m/s). We carried out the calculations for different
implementations of sequences of random numbers (see Section 3) and
different numbers of atoms in the ensemble. These allow us to estimate
the accuracy of the calculations. Comparison of the curves in
Fig.~\ref{fig-Cs-v-a-dv},~\emph{a} allows us to talk about the
accuracy of calculations of $a$ at the level of 7--10\,\% for 500
atoms and up to 20\,\% for 100 atoms. Comparison of the curves in
Fig.~\ref{fig-Cs-v-a-dv},~\emph{b} allows us to estimate the accuracy
of the calculations of $\Delta{}v$ at the level 15\,\% for 500 atoms
and 25\,\% for 100 atoms.

Since the velocity of atoms changes continuously with time, the
momentum diffusion coefficient also changes with time.  To completely
exclude the effect of a change in velocity on the momentum diffusion
coefficient, in the next section we consider the change in momentum of
an atom with time for the ``heavy atom'' model when the change in
atomic velocity can be neglected. In addition, this consideration will
allow us to compare the results of the calculation of the light
pressure force exerted on an atom in the quasiclassical
theory~\cite{Voitsekhovich-88,Yatsenko,Podlecki:18} with the results
of the quantum mechanical theory. Since in quasi-classical theory the
atom is considered as a material point, and in our calculations, the
atom at the beginning of interaction with the field is considered as a
monochromatic wave, consistency of the results obtained from these
opposite assumptions is important for the confident application of the
quasi-classical approach in the cases where we are interested only in
the light pressure force exerted on an atom.

\subsection[Важкі атоми у біхроматичному полі. Сила світлового тиску і
дисперсія імпульсів]{Heavy atoms in the field of counter-propagating
  bichromatic waves. Light pressure force and momentum variance}
\label{sec:quant-heavy}

\begin{figure}[h]
\includegraphics[width=86mm]{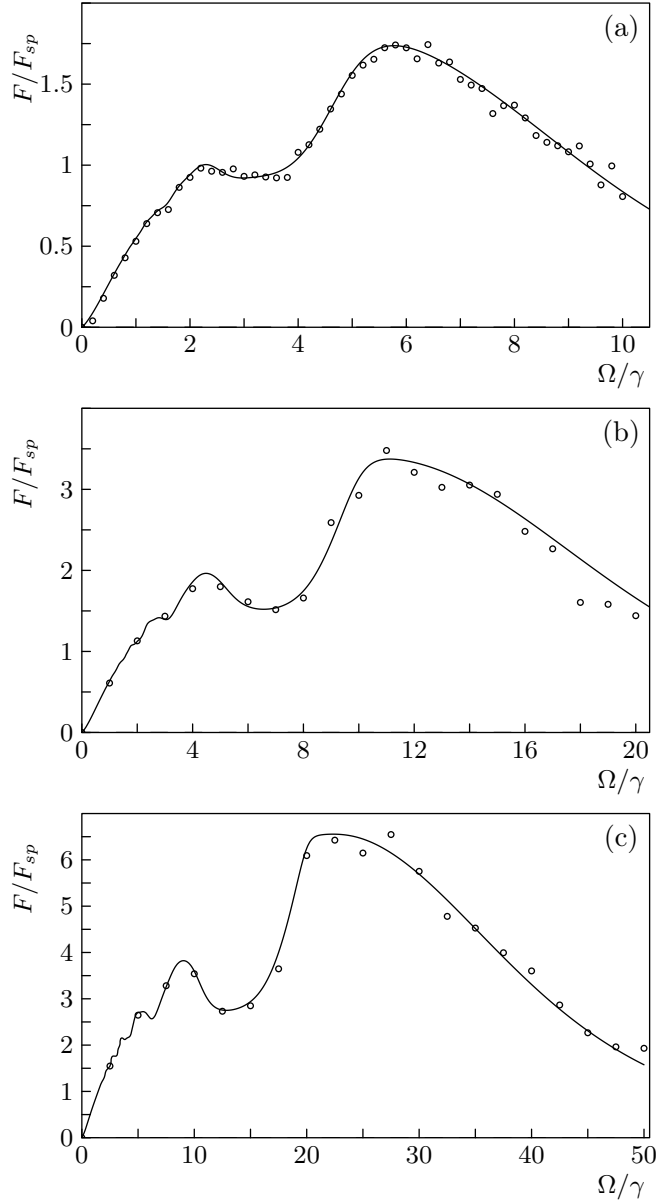}
  \caption{The dependence of the light pressure force
    exerted on an atom in units of $F_{sp}$ on $\Omega/\gamma$. Solid
    curves --- calculation based on the results of
    work~\cite{Voitsekhovich-88}, circles ~ --- calculation by the
    Monte Carlo wave function method. (a) ~ ---
    $\Omega_{0}/\gamma=$6.124, 1000 atoms; (b) ~ ---
    $\Omega_{0}/\gamma=$12.25, 100 atoms; (c) ~ ---
    $\Omega_{0}/\gamma=$24.5, 1000 atoms. The initial velocity of the
    atoms is $v_{0}=0$. Phase difference of the counter-propagating
    waves is $\varphi=\pi/4$}
\label{fig-R-inf-F}
\end{figure}

We simulate the motion of ``heavy'' atoms in the field of the
counter-propagating waves by the procedure described in
Sec.~\ref{sec:numerical} and use Eqs.~(\ref{eq:bg-inf}),
(\ref{eq:be-inf}), which describe the time evolution of the
probability amplitudes $b_{g,n}$, $b_{e,n}$.  Calculating the change
of the average momentum of the atoms during the time
$\sim100\gamma^{-1}$, much greater than the time of the transient
processes at the beginning of the interaction of atoms with the field
($\sim10\gamma^{-1}$), we find the average force exerted on an atom.
In addition, we calculate the light pressure force by the density
matrix method according to the theoretical
work~\cite{Voitsekhovich-88}. As can be seen from
Fig.~\ref{fig-R-inf-F}, the calculations by both methods are quite
close, despite the relatively low accuracy of the calculation by the
Monte Carlo method for an ensemble of 100--1000 atoms.

\begin{figure}[h]
\includegraphics[width=86mm]{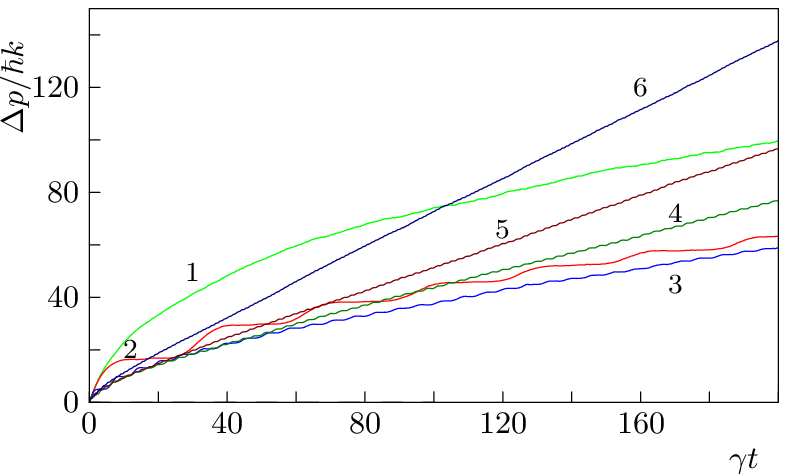}
 \caption{Time dependence of the standard deviation
$\Delta{}p$ of the atom's momentum from its average value for for 1000
atoms and $\Omega_{0}/\gamma=6.124$. Parameters are: 1~-- $\Omega=0$;
2~-- $\Omega=0.2\gamma$; 3~-- $\Omega=\gamma$; 4~-- $\Omega=2\gamma$,
5~-- $\Omega=3\gamma$, 6~-- $\Omega=5\gamma$. The initial velocity of
the atoms is $v_{0}=0$. Phase difference of the counter-propagating
waves is $\varphi=\pi/4$}
\label{fig-inf-dpO}
\end{figure}

Now let's consider the change of the standard deviation $\Delta{}p$ of
the momentum from its average value (square root of the variance of
the momentum) on time. In the previous section, it was shown that at
the beginning of the interaction of the atom with the field
$\Delta p\sim\sqrt{t}$ due to momentum diffusion. Further, because of
the change of the atomic velocity, the law describing the dependence
of $\Delta{}p$ on time changes. Here we consider similar dependencies
for the ``heavy atom'' model. An example of the numerical simulations
is shown in Fig.~\ref{fig-inf-dpO} for zero initial velocity of atoms.
As we can see, for a standing monochromatic wave (curve 1)
$\Delta{}p\sim{}\sqrt{t}$, which indicates the diffusion nature of the
blurring of the distribution of atoms by momentum. The momentum
diffusion coefficient is $D=74.3D_{r}$. Even a relatively small
difference of frequencies between the components of bichromatic waves,
$\Omega=0.2\gamma$, leads to a significant change in the dependence of
$\Delta{}p(t)$. As can be seen, for the initial moments of time,
$t<4\gamma^{-1}$, curves 1 and 2 almost coincide, later on, curve 2
oscillations are observed, the period of which is $2\pi/\Omega$. The
momentum diffusion coefficient determined from the approximation of
curve 2 by the root dependence is equal to $D=27.6D_{r}$, i.e. 2.7
times less than for a standing monochromatic wave (curve 1).  As
$\Omega$ increases, the amplitude of the oscillations and their period
decrease (curves 3--6), and the root dependence of smoothed curves
3--6 of $\Delta{}p $ on time becomes almost linear. Curve 2 is similar
to the root dependence (taking into account the smoothing of
oscillations), curve 3 at the beginning is also close to the graph of
the square root, but closer to the end of the abscissa axis is more
like a graph of a linear function, and curve 6 is a straight line on
the almost whole abscissa axis interval. Thus, for small values of
$\Omega$ we have the diffusion of atoms in the momentum space, and for
greater $\Omega$ the dependence $\Delta{}p(t)$ corresponds to the
scattering of atoms in a certain range of angles. In addition, with
the transition from a monochromatic field to a bichromatic one, even
at $\Omega\ll\gamma$ the momentum diffusion coefficient decreases
sharply.

Comparing Fig.~\ref{fig-inf-dpO} and Fig.~\ref{fig-R122-dv}, we see an
obvious contradiction: in Fig.~\ref{fig-R122-dv} the diffusion-like
dependence of the standard deviation $\Delta{}v$ of the velocity of
atoms on its average value for $\Omega\sim\Omega_{0}$ is clearly
visible, at the same time in Fig.~\ref{fig-inf-dpO} analogous
dependence for the standard deviation of the atom's momentum from its
average value is close to linear. Since the dependences shown in
Fig.~\ref {fig-R122-dv} are calculated taking into account the change
in the velocity of atoms over time, we should investigate the
dependence of $\Delta{}p $ on time for a nonzero initial velocity of a
``heavy'' atom.

\begin{figure}[th]
\includegraphics[width=86mm]{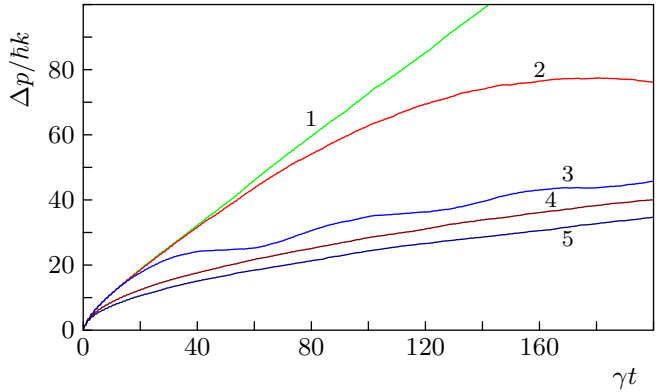}
\caption{Time dependence of the standard deviation
    $\Delta{}p$ of the atom's momentum from its average value for 1000
    atoms and $\Omega_{0}=6.124\gamma$, $\Omega=5\gamma$. Parameters
    are: 1~-- $kv_{0}=0$, 2~-- $kv_{0}=0.01\gamma$, 3~--
    $kv_{0}=0.05\gamma$, 4~-- $kv_{0}=\gamma$, 5~--
    $kv_{0}=2\gamma$. Phase difference of the counter-propagating
    waves is $\varphi=\pi/4$}
\label{fig-inf-dpt-kv}
\end{figure}

Fig.~\ref{fig-inf-dpt-kv} shows the time dependence of the standard
deviation $\Delta{}p$ of the atom's momentum from its average value
for $\Omega_{0}=6.124\gamma$, $\Omega=5\gamma$ (optimal ratio of
$\Omega$ and $\Omega_ {0}$ according to~\cite{Yatsenko}).
As can be seen, the case $v_{0}=0$ is special: only in this case the
dependence of $\Delta{}p$ on $t$ is linear for $\gamma{}t>10$. If the
initial velocity is nonzero, oscillations with frequency $kv_{0}$
appear on the specified dependence. The dependence $\Delta{}p(t)$ for
$v_{0}=0$ can be considered as oscillating with zero frequency. At the
beginning of the interaction with the field, at $\gamma{}t\ll1$, all
the graphs almost coincide with the curve constructed for
$v_{0}=0$. As can be seen, for $kv_{0}\ge\gamma$, besides the initial
moments of time, when the transients associated with the spontaneous
emission play a significant role, the dependence of $\Delta{}p$ on
time is described by the law of quadratic root. Thus, for real atoms
of finite mass, in which the velocity changes during interaction with
the field, momentum diffusion should be observed, as shown in
Fig.~\ref{fig-R122-dv}.

Momentum diffusion coefficients, found by approximating the curves
shown in Fig.~\ref{fig-inf-dpt-kv} with the square root law using the
method of least squares, are: $D=14.9D_{r}$ (curve 3), $D = 11.6D_{r}$
(curve 4), $D=8.4D_{r}$ (curve 5).

\begin{figure}[h!]
  \includegraphics[width=86mm]{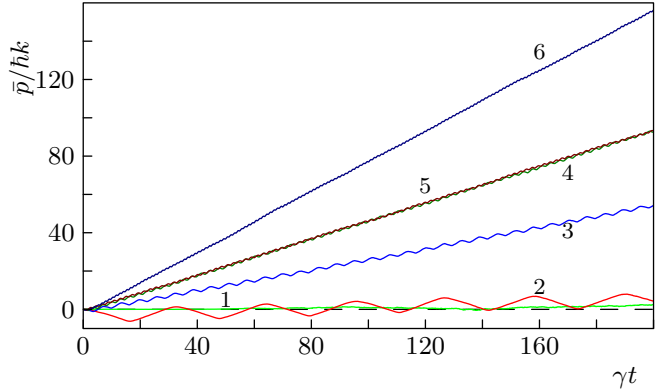}
  \caption{ Time dependence of the average momentum
    $\bar{p}$ of an atom for 1000 atoms and
    $\Omega_{0}/\gamma=6.124$. Parameters are: 1~-- $\Omega = 0 $,
    2~-- $\Omega = 0.2\gamma$, 3~-- $\Omega = \gamma$, 4~--
    $\Omega = 2\gamma$, 5~-- $\Omega = 3\gamma$, 6~--
    $\Omega = 5\gamma$. The initial velocity of the atoms $v_{0} =
    0$. Phase difference of counter waves $\varphi = \pi/4$}
\label{fig-inf-pO}
\end{figure}

Oscillations with a period corresponding to the frequency difference
between the monochromatic components of the bichromatic waves are also
observed in the corresponding to Fig.~\ref{fig-inf-dpO} time
dependencies of the average momentum $\bar{p}$ of atoms (see
Fig.~\ref{fig-inf-pO}).
As can be seen from Fig.~\ref{fig-inf-pO}, curves 4 and 5 almost
coincide, which indicates the same light pressure force exerted on the
atoms for $\Omega=2\gamma$ and $\Omega=3\gamma$, in conformity with
Fig.~\ref{fig-R-inf-F}, \emph{a}.

Fig.~\ref{fig-inf-dpt-kv} shows that the width of the atomic momentum
distribution at the beginning of the atom-field interaction for zero
initial velocity of atoms linearly depends on time. When applying the
calculations to real atoms, it should be borne in mind that even a
slight change in the velocity of atoms is enough to change the linear
law to nonlinear, which is well described by the law of the square
root at $kv_{0}=0.05\gamma$.

\begin{figure}[h!]
\includegraphics[width=86mm]{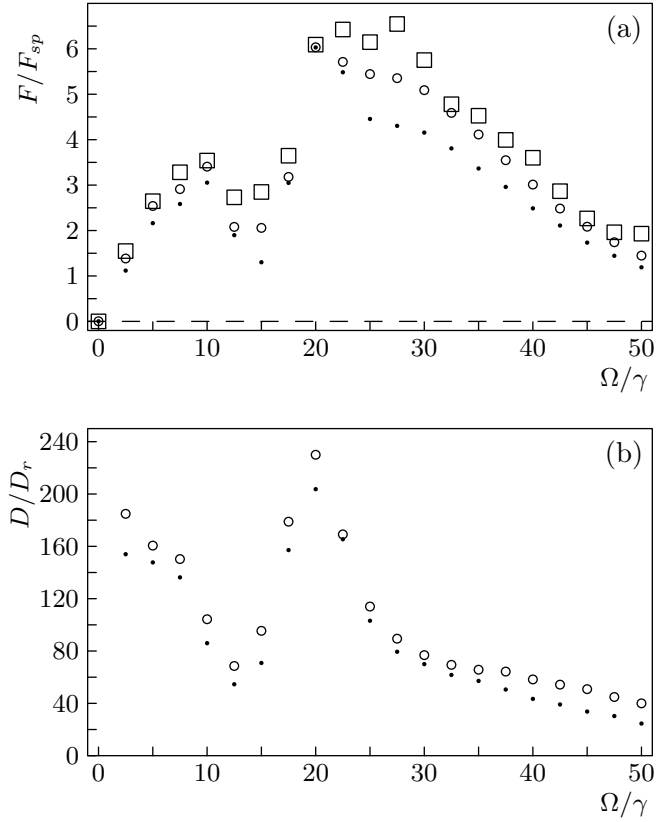}
  \caption{ Dependencies of the light pressure force $F$ in
    units of $F_{sp}$ (\emph{a}) and the momentum diffusion
    coefficient (\emph{b}) on $\Omega/\gamma$ for different initial
    velocities $v_{0}$ of atoms for
    $\Omega_{0}/\gamma=24.5$. Averaging per 1000 atoms, the phase
    difference of the counter-propagating waves is
    $\varphi=\pi/4$. Rings~-- $kv_{0}/\gamma=0.1$, circles~--
    $ kv_{0}/\gamma=0.2$, squares~-- $v_{0}=0$. For the case
    $\Omega=0$ we have (not shown) $D/D_r=1371$ for
    $kv_{0}/\gamma=0.1$ and $D/D_r=1259$ for $kv_{0}/\gamma=0.2$}
\label{fig-inf-F-D-Omega}
\end{figure}

The dependencies of the momentum diffusion coefficient and the average
light pressure force on $\Omega$ for $kv_{0}=0.1\gamma{}$ and
$kv_{0}=0.2\gamma{}$ are shown in Fig.~\ref{fig-inf-F-D-Omega}.
As can be seen in Fig.~\ref{fig-inf-F-D-Omega}, \emph{a}, for
$\Omega=20\gamma=\sqrt{2/3}\Omega_{0}$ the light pressure force for
both values of $ kv_{0}$ is almost the same. According to the
work~\cite{Yatsenko}, this should be the case with the optimal ratio
of $\Omega_{0}/\Omega=\sqrt{3/2}$, when the light pressure force
changes a little with the speed of the atom near the maximum value of
force. The momentum diffusion coefficient as a function of $\Omega$
(see Fig.~\ref{fig-inf-F-D-Omega}, \emph{b}) reaches a local maximum
at this point. The absolute maximum of the momentum diffusion
coefficient ($D/D_r=1371$, $D/D_r=1259$ for $kv_ {0}/\gamma=0.1$ and
$kv_{0}/\gamma=0.2$ respectively) is reached only at $\Omega=0 $ when
the counter-propagating bichromatic waves become monochromatic. Thus,
the momentum diffusion coefficient under the condition of the optimal
ratio of $\Omega$ and $\Omega_{0}$ is six times smaller than the
momentum diffusion coefficient in the counter-propagating
monochromatic waves, in which the counter-propagating bichromatic
waves degenerate at $\Omega=0$.

\begin{figure}[h!]
\includegraphics[width=86mm]{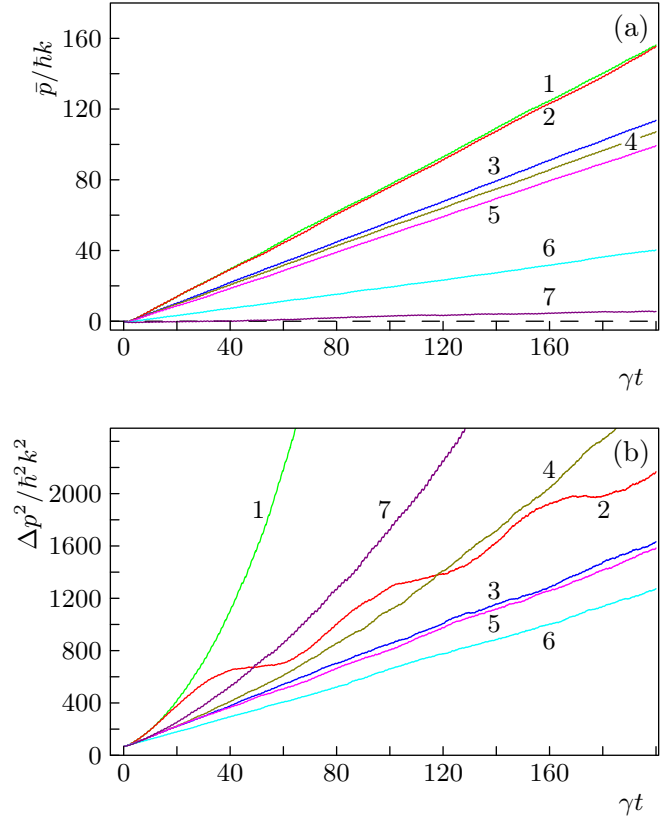}
\caption{Time dependencies of the average momentum $\bar{p}$ of atoms
  (\emph{a}) and the average square of the deviation of the momentum
  from the average value $\Delta{}p^{2}$ (\emph{b}). Parameters are:
  $\Omega_{0}/\gamma=6.124$, $\Omega/\gamma=5$. Averaging per 1000
  atoms. The phase difference of counter waves is
  $\varphi=\pi/4$. 1~-- $v_{0}=0$, 2~-- $kv_{0}/\gamma=0.05$, 3~--
  $kv_{0}/\gamma=1.2$, 4~-- $kv_{0}/\gamma=1.25$, 5~--
  $kv_{0}/\gamma=1.3$, 6~-- $kv {0}/\gamma=2$, 7~--
  $ kv_{0}/\gamma=2.5$}
\label{fig-inf-F-D2}
\end{figure}

Besides the case of zero initial velocity of atoms (recall that in the
approximation of a heavy atom we neglect the change of velocity,
assuming that only the momentum of the atom changes with time), there
are other values of the initial velocity of atoms at which the
dependence $\Delta{}p(t)$ is quite well described by linear law. The
transition from scattering to diffusion and vice versa with a change
in the initial velocity of the atom is illustrated in
Fig.~\ref{fig-inf-F-D2}, It shows the time dependence of
$\Delta{}p^{2}(t)$, which in the case of momentum diffusion is close
to linear. Fig.~\ref{fig-inf-F-D2} also shows the time dependence of
the varying part $\bar{p} $ of the total average momentum of atoms
$mv_{0}+\bar{p}$, which we will call as ``the average momentum of the
atom'' to simplify the terminology.
It is noteworthy that while curves 1 and 2 in
Fig.~\ref{fig-inf-F-D2},~\emph{a} are very close, the corresponding
curves in Fig.~\ref{fig-inf-F-D2}, \emph{b} are close only at the
beginning of the interaction of the atoms with the field. Curve 1 in
Fig.~\ref{fig-inf-F-D2}, \emph{b} is a parabola, it corresponds to the
linear change of $\Delta{p}$ with time. Thus, in the case of $v_{0}=0$
we have the scattering of atoms. Even when the initial velocity of
atoms becomes relatively small (see curve 2 in
Fig.~\ref{fig-inf-F-D2}, \emph{b}), the time dependence of
$\Delta{p}^{2}$ dramatically changes. It becomes almost, with accuracy
to small oscillations, a linear function, and $\Delta{}(t)$ is
approximately described by the function $\Delta{}p=\sqrt{2Dt}$. In
this case, we can talk about the momentum diffusion, which is
described by the momentum diffusion coefficient $D$. As the initial
velocity increases, $\Delta{}p(t)$ is approximately described by the
diffusion law until the velocity becomes $v_{0}=\Omega/(4k)$. We see
that the time dependence of the average momentum of atoms at
$v_{0}=\Omega/(4k)$ (curve~4) differs a little from similar
dependences at a small change in velocity in one direction or another
(curves 3 and 5); the corresponding dependencies of $\Delta{}p^{2}$ on
time differ radically: curve 4 is a parabola (scattering of atoms),
curves 3 and 5~-- are graphs of linear functions (momentum
diffusion). Probably the transition to scattering at the initial
velocity of atoms equals $v_{0}=\Omega/(4k)$ is associated with
Doppleron resonance \cite{Freund,Kyrola77}, which in this case is due
to a four-photon process: the absorption of two quanta of light from
the monochromatic component of the counter-propagating bichromatic
wave of frequencies $\omega+\frac{1}{2}\Omega+kv_{0}$ and
$\omega-\frac{1}{2}\Omega+kv_{0}$ and the emission of two quanta of
light into a monochromatic component of a concomitant bichromatic wave
of frequency $\omega+\frac{1}{2}\Omega-kv_{0}$ or a four-photon
process with the absorption of two quanta from the monochromatic
component of frequency $\omega-\frac{1}{2}\Omega+kv_{0}$ from the
counter-propagating wave and the emission of quanta with frequencies
$\omega-\frac{1}{2}\Omega-kv_{0}$, $\omega+\frac{1}{2}\Omega-kv_{0}$
in monochromatic components of the concomitant bichromatic wave. With
increasing $v_{0}$ we again fall into the region of momentum diffusion
(see curve 6). Other theoretically possible Doppleron resonances (for
example, at $v_{0}=\Omega/(3k)$) at the parameters of interaction of
atoms with the field corresponding to Fig.~\ref{fig-inf-F-D2} were not
registred, until reaching the velocity $v_{0}=\Omega/(2k)$ (curve 7),
where the time dependence of $\Delta{}p$ corresponds to the scattering
of atoms. Note that this velocity corresponds to the single-photon
resonance of waves with frequencies $\omega-\frac{1}{2}\Omega+kv_{0}$
and $\omega+\frac{1}{2}\Omega-kv_{0}$ and reducing the force of light
pressure almost to zero \cite{Voitsekhovich-88}.

\begin{figure}[t]
\includegraphics[width=86mm]{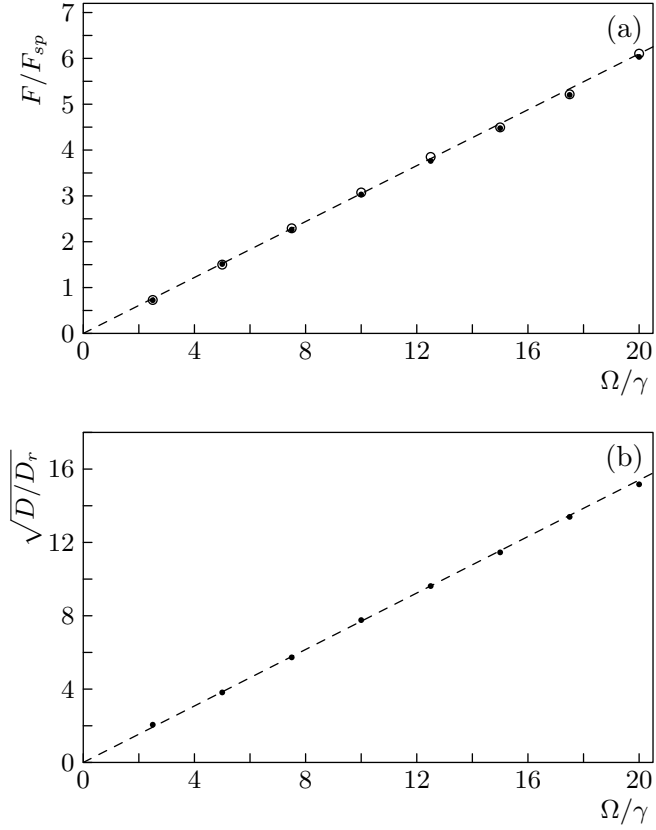}
  \caption{ Dependence of the light pressure force exerted
    on atoms in units of $F_{sp}$ (\emph{a}) and the square root of the
    coefficient of momentum diffusion $\sqrt{D}$ in units of
    $\sqrt{D_{r}} $ (\emph{b}) on the frequency difference of the
    components of the counter-propagating bichromatic wave in units of
    $\gamma$ provided that the Rabi frequency
    $\Omega_{0}=\sqrt{3/2}\Omega$ and $\varphi=\pi/4$. Averaging per
    1000 atoms. Rings correspond to $v_{0}=0$, circles correspond to
    $kv_{0}=0.1\gamma$}
\label{fig-inf-F-D-Omega-opt}
\end{figure}

Fig.~\ref{fig-inf-F-D-Omega-opt} shows the dependence of the light
pressure force $F$ exerted on atoms in units of $F_{sp}$ and the
square root of the momentum diffusion coefficient $\sqrt{D}$ in units
of $\sqrt{D_{r}}$ on the difference of frequencies of the
monochromatic components of the counter-propagating bichromatic waves
in units of $\gamma$ provided that the optimal conditions of the
atom-field interaction are fulfilled: $\Omega_{0}=\sqrt{3/2}\Omega$,
$\varphi=\pi/4$.
The values of the light pressure force $F=\bar{p}/t$ and the pulse
diffusion coefficient $D=\Delta{}p^{2}/(2t) $ were calculated by the
method of least squares from the dependencies $\bar{p}(t)$ and
$\Delta{}p^{2}(t)$ which were calculated for an ensemble of 1000 atoms
in the time interval from 0 to $100/\gamma$. Using the dependence
shown in Fig.~\ref{fig-inf-F-D-Omega-opt}, \emph{b}, we find for the
case $kv_{0}=0.1\gamma$:
\begin{equation}
  \label{eq:Dcalc}
  D\approx{}0.6D_{r}\frac{\Omega^{2}}{\gamma^{2}}.
\end{equation}
This is about twice as much, as gives the Eq.~\eqref{eq:Dbi}.

As expected, according to the theoretical work~\cite{Yatsenko}, the
force exerted on an atom under optimal conditions of the atom-field
interaction depends linearly on $\Omega$. According to our
calculations, the square root of the pulse diffusion coefficient also
depends linearly on $\Omega$. This means that the ratio $\sqrt{D}/F$,
and, as a result
\begin{equation}
  \label{eq:dpp}
  \frac{\Delta{}p}{\bar{p}}=\frac{\sqrt{2Dt}}{Ft}
 \approx2.5\frac{\sqrt{2D_{r}}}{F_{sp}\sqrt{t}}
  \approx\frac{4.2}{\sqrt{\gamma{}t}}
\end{equation}
does not depend on $\Omega$.  Since
$\Omega_{0}=\sqrt{\frac{3}{2}}\Omega$, the momentum diffusion
coefficient under optimal conditions of the atom-field interaction is
proportional to $\Omega_{0}^{2}$, i.e. the intensity of the laser
radiation.

\begin{figure}[t]
  \includegraphics[width=86mm]{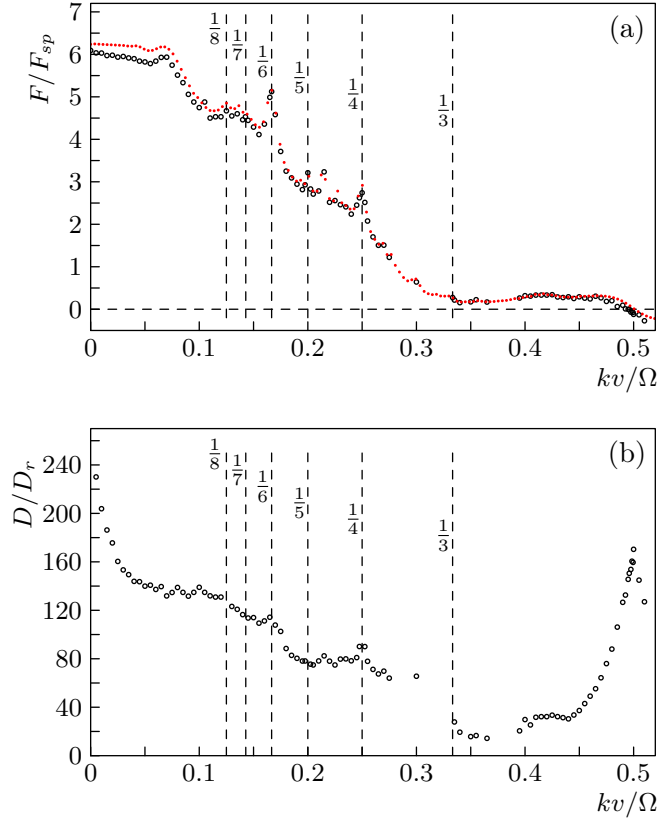}
  \vskip-3mm\caption{ The dependences of the light pressure force $F$
    exerted on atoms in units of $F_{sp}$ (\emph{a}) and the momentum
    diffusion coefficient ${D}$ in units of ${D_{r}}$ (\emph{b}) on
    the atomic velocity for $\Omega_{0}=\sqrt{3/2}\Omega$,
    $\Omega=20\gamma$ and $\varphi=\pi/4$. Averaging per 1000
    atoms. Rings are the results of calculation by the Monte Carlo
    wave function method, the solid curve is the results of
    calculations by the density matrix equation}
\label{fig-inf-F-D-kv}
\end{figure}

Fig.~\ref{fig-inf-F-D-kv} shows the dependence of the light pressure
force exerted on atoms in the field of counter-propagating bichromatic
waves and the momentum diffusion coefficient ${D}$ on the atomic
velocity for the optimal ratio $\Omega_{0}/\Omega=\sqrt{3/2}$ and
$\varphi=\pi/4$, which were found from the time dependence of the
change of the average momentum and the average square momentum during
time $t=100/\gamma$. The results of the calculation of the light
pressure force on atoms by the equations for the density matrix,
similar to the works~\cite{Voitsekhovich-88,Soding}, are also
presented. The dependencies of the force on velocity, calculated by
both methods, almost coincide. Some differences can be explained by
different descriptions of the atom: in our calculations, the atom at
the beginning of the interaction with the field is a plane wave, in
the calculations by the equations for the density matrix the atom is a
material point. It is noteworthy that the coefficient of momentum
diffusion with increasing velocity $v$ of atoms at $kv<0.05\gamma$
decreases rather quickly, almost twice when velocity changes from
$v=0.005\Omega/k$ to $v=0.05\Omega/k$, while the light pressure force
remains almost unchanged in this velocity range.  Note that points on
the graph correspond to the velocity $v$ of different ensembles of
atoms, which for each ensemble in the approximation of ``heavy'' atom
during its movement remains unchanged and equal to the initial
velocity $v=v_{0}$. At velocities corresponding to Doppleron
resonances (denoted by vertical dotted lines), the momentum diffusion
coefficient could not be calculated because the time dependence of the
variance of the momentum is not described by the law
$\Delta{}p^{2}={2Dt}$.

The results of this section are valid in the approximation of a heavy
atom when the change of its velocity during the atom-field interaction
is neglected. For real atoms, the transition from scattering to
momentum diffusion and vice versa occurs in a short time and is not
noticeable in Fig.~\ref{fig-R122-v} and Fig.~\ref{fig-R122-dv},
perhaps due to insufficient accuracy of our calculations by the Monte
Carlo method.

\section{Conclusions}
\label{sec:conclusion}

The calculation of the light pressure force exerted on atoms in the
field of counter-propagating bichromatic waves based on the quantum
mechanical description of the mechanical motion of atoms agrees well
with the previous studies of the light pressure based on the classical
description of mechanical motion. That is an additional substantiation
of the classical approach to the problems of the motion of atoms in
the field of the counter-propagating modulated waves.

The standard deviation of the velocity from its average value may
increase or decrease with time. This correlates with the increase or
decrease of the average acceleration of atoms and is associated with
both the momentum diffusion of atoms and bunching or the antibunching
of atoms around the average velocity of the atomic ensemble.

We calculated the momentum diffusion of atoms in the field of
counter-propagating bichromatic waves both for ideal cases of
``heavy'' atoms and for sodium atoms.  We have shown that the analogy
between the interaction of atoms with the field of counter-propagating
bichromatic waves and the interaction of atoms with the field of
counter-propagating $\pi$-pulses can be used for a rough estimation of
the momentum diffusion coefficient of the atoms in the field of the
counter-propagating bichromatic waves.

Depending on the parameters of the interaction of atoms with the
field, the time dependence of the momentum variance corresponds to
either scattering of atoms (square root of the momentum variance is
proportional to time) or diffusion of atoms (the momentum variance is
proportional to time). Momentum diffusion changes to the scattering of
atoms when the velocity of atoms approaches Doppleron resonances.

It is shown that under the conditions of optimal parameters of the
interaction of atoms with the field, the coefficient of momentum
diffusion is proportional to the intensity of laser radiation. Under
these conditions, the ratio of the root mean square deviation of
atomic momentum from its average value to the average atomic momentum
does not depend on $\Omega$.

The publication contains the results of research in the frame of
grant support of the target complex program of fundamental research of
NAS of Ukraine ``Fundamental problems of creation of new
nanomaterials and nanotechnologies'', contract No 3/19-N, as well
as on the theme V-185.


\begin{thebibliography}{20}
\expandafter\ifx\csname natexlab\endcsname\relax\def\natexlab#1{#1}\fi
\expandafter\ifx\csname bibnamefont\endcsname\relax
  \def\bibnamefont#1{#1}\fi
\expandafter\ifx\csname bibfnamefont\endcsname\relax
  \def\bibfnamefont#1{#1}\fi
\expandafter\ifx\csname citenamefont\endcsname\relax
  \def\citenamefont#1{#1}\fi
\expandafter\ifx\csname url\endcsname\relax
  \def\url#1{\texttt{#1}}\fi
\expandafter\ifx\csname urlprefix\endcsname\relax\def\urlprefix{URL }\fi
\providecommand{\bibinfo}[2]{#2}
\providecommand{\eprint}[2][]{\url{#2}}

\bibitem[{\citenamefont{Voitsekhovich et~al.}(1988)\citenamefont{Voitsekhovich,
  Danileiko, Negriiko, Romanenko, and Yatsenko}}]{Voitsekhovich-88}
\bibinfo{author}{\bibfnamefont{V.~S.} \bibnamefont{Voitsekhovich}},
  \bibinfo{author}{\bibfnamefont{M.~V.} \bibnamefont{Danileiko}},
  \bibinfo{author}{\bibfnamefont{A.~M.} \bibnamefont{Negriiko}},
  \bibinfo{author}{\bibfnamefont{V.~I.} \bibnamefont{Romanenko}},
  \bibnamefont{and} \bibinfo{author}{\bibfnamefont{L.~P.}
  \bibnamefont{Yatsenko}}, \bibinfo{journal}{Sov. Phys. Tech. Phys.}
  \textbf{\bibinfo{volume}{33}}, \bibinfo{pages}{690} (\bibinfo{year}{1988}).

\bibitem[{\citenamefont{Minogin and Letokhov}(1987)}]{Minogin}
\bibinfo{author}{\bibfnamefont{V.~G.} \bibnamefont{Minogin}} \bibnamefont{and}
  \bibinfo{author}{\bibfnamefont{V.~S.} \bibnamefont{Letokhov}},
  \emph{\bibinfo{title}{Laser Light Pressure on Atoms}}
  (\bibinfo{publisher}{Gordon and Breach: New York}, \bibinfo{year}{1987}).

\bibitem[{\citenamefont{Metcalf and van~der Stratten}(1999)}]{Metcalf-99}
\bibinfo{author}{\bibfnamefont{H.~J.} \bibnamefont{Metcalf}} \bibnamefont{and}
  \bibinfo{author}{\bibfnamefont{P.}~\bibnamefont{van~der Stratten}},
  \emph{\bibinfo{title}{Laser Cooling and Trapping}}
  (\bibinfo{publisher}{Springer-Verlag: New York, Berlin, Heidelberg},
  \bibinfo{year}{1999}).

\bibitem[{\citenamefont{Vo\u{\i}tsekhovich
  et~al.}(1989)\citenamefont{Vo\u{\i}tsekhovich, Danile\u{\i}ko, Negri\u{\i}ko,
  Romanenko, and Yatsenko}}]{Voitsekhovich-89}
\bibinfo{author}{\bibfnamefont{V.~S.} \bibnamefont{Vo\u{\i}tsekhovich}},
  \bibinfo{author}{\bibfnamefont{M.~V.} \bibnamefont{Danile\u{\i}ko}},
  \bibinfo{author}{\bibfnamefont{A.~N.} \bibnamefont{Negri\u{\i}ko}},
  \bibinfo{author}{\bibfnamefont{V.~I.} \bibnamefont{Romanenko}},
  \bibnamefont{and} \bibinfo{author}{\bibfnamefont{L.~P.}
  \bibnamefont{Yatsenko}}, \bibinfo{journal}{JETP Lett.}
  \textbf{\bibinfo{volume}{49}}, \bibinfo{pages}{161} (\bibinfo{year}{1989}).

\bibitem[{\citenamefont{S\"oding et~al.}(1997)\citenamefont{S\"oding, Grimm,
  Ovchinnikov, Bouyer, and Salomon}}]{Soding}
\bibinfo{author}{\bibfnamefont{J.}~\bibnamefont{S\"oding}},
  \bibinfo{author}{\bibfnamefont{R.}~\bibnamefont{Grimm}},
  \bibinfo{author}{\bibfnamefont{Y.}~\bibnamefont{Ovchinnikov}},
  \bibinfo{author}{\bibfnamefont{P.}~\bibnamefont{Bouyer}}, \bibnamefont{and}
  \bibinfo{author}{\bibfnamefont{C.}~\bibnamefont{Salomon}},
  \bibinfo{journal}{Phys.\ Rev.\ Lett.} \textbf{\bibinfo{volume}{78}},
  \bibinfo{pages}{1420} (\bibinfo{year}{1997}).

\bibitem[{\citenamefont{Yatsenko and Metcalf}(2004)}]{Yatsenko}
\bibinfo{author}{\bibfnamefont{L.}~\bibnamefont{Yatsenko}} \bibnamefont{and}
  \bibinfo{author}{\bibfnamefont{H.}~\bibnamefont{Metcalf}},
  \bibinfo{journal}{Phys.\ Rev.\ A} \textbf{\bibinfo{volume}{70}},
  \bibinfo{pages}{063402} (\bibinfo{year}{2004}).

\bibitem[{\citenamefont{Podlecki et~al.}(2018)\citenamefont{Podlecki, Glover,
  Martin, and Bastin}}]{Podlecki:18}
\bibinfo{author}{\bibfnamefont{L.}~\bibnamefont{Podlecki}},
  \bibinfo{author}{\bibfnamefont{R.~D.} \bibnamefont{Glover}},
  \bibinfo{author}{\bibfnamefont{J.}~\bibnamefont{Martin}}, \bibnamefont{and}
  \bibinfo{author}{\bibfnamefont{T.}~\bibnamefont{Bastin}},
  \bibinfo{journal}{J. Opt. Soc. Am. B} \textbf{\bibinfo{volume}{35}},
  \bibinfo{pages}{127} (\bibinfo{year}{2018}).

\bibitem[{\citenamefont{Metcalf}(2017)}]{Metcalf}
\bibinfo{author}{\bibfnamefont{H.}~\bibnamefont{Metcalf}},
  \bibinfo{journal}{Rev. Mod. Phys.} \textbf{\bibinfo{volume}{89}},
  \bibinfo{pages}{041001} (\bibinfo{year}{2017}).

\bibitem[{\citenamefont{Berg-S{\o}renson
  et~al.}(1992)\citenamefont{Berg-S{\o}renson, Castin, Bonderup, and
  M{\o}lmer}}]{Berg-Sorenson}
\bibinfo{author}{\bibfnamefont{K.}~\bibnamefont{Berg-S{\o}renson}},
  \bibinfo{author}{\bibfnamefont{Y.}~\bibnamefont{Castin}},
  \bibinfo{author}{\bibfnamefont{E.}~\bibnamefont{Bonderup}}, \bibnamefont{and}
  \bibinfo{author}{\bibfnamefont{K.}~\bibnamefont{M{\o}lmer}},
  \bibinfo{journal}{Journal of Physics B: Atomic, Molecular and Optical
  Physics} \textbf{\bibinfo{volume}{25}}, \bibinfo{pages}{4195}
  (\bibinfo{year}{1992}).

\bibitem[{\citenamefont{Vo\u{\i}tsekhovich
  et~al.}(1991)\citenamefont{Vo\u{\i}tsekhovich, Danile\u{\i}ko, Negri\u{\i}ko,
  Romanenko, and Yatsenko}}]{JETP}
\bibinfo{author}{\bibfnamefont{V.~S.} \bibnamefont{Vo\u{\i}tsekhovich}},
  \bibinfo{author}{\bibfnamefont{M.~V.} \bibnamefont{Danile\u{\i}ko}},
  \bibinfo{author}{\bibfnamefont{A.~N.} \bibnamefont{Negri\u{\i}ko}},
  \bibinfo{author}{\bibfnamefont{V.~I.} \bibnamefont{Romanenko}},
  \bibnamefont{and} \bibinfo{author}{\bibfnamefont{L.~P.}
  \bibnamefont{Yatsenko}}, \bibinfo{journal}{Sov. Phys. JETP}
  \textbf{\bibinfo{volume}{72}}, \bibinfo{pages}{219} (\bibinfo{year}{1991}).

\bibitem[{\citenamefont{Dalibard et~al.}(1992)\citenamefont{Dalibard, Castin,
  and M{\o}lmer}}]{Dalibard}
\bibinfo{author}{\bibfnamefont{J.}~\bibnamefont{Dalibard}},
  \bibinfo{author}{\bibfnamefont{Y.}~\bibnamefont{Castin}}, \bibnamefont{and}
  \bibinfo{author}{\bibfnamefont{K.}~\bibnamefont{M{\o}lmer}},
  \bibinfo{journal}{Phys.\ Rev.\ Lett.} \textbf{\bibinfo{volume}{68}},
  \bibinfo{pages}{580} (\bibinfo{year}{1992}).

\bibitem[{\citenamefont{M{\o}lmer et~al.}(1993)\citenamefont{M{\o}lmer, Castin,
  and Dalibard}}]{Molmer}
\bibinfo{author}{\bibfnamefont{K.}~\bibnamefont{M{\o}lmer}},
  \bibinfo{author}{\bibfnamefont{Y.}~\bibnamefont{Castin}}, \bibnamefont{and}
  \bibinfo{author}{\bibfnamefont{J.}~\bibnamefont{Dalibard}},
  \bibinfo{journal}{JOSA B} \textbf{\bibinfo{volume}{10}}, \bibinfo{pages}{524}
  (\bibinfo{year}{1993}).

\bibitem[{\citenamefont{Corder et~al.}(2015{\natexlab{a}})\citenamefont{Corder,
  Arnold, Hua, and Metcalf}}]{Corder:15}
\bibinfo{author}{\bibfnamefont{C.}~\bibnamefont{Corder}},
  \bibinfo{author}{\bibfnamefont{B.}~\bibnamefont{Arnold}},
  \bibinfo{author}{\bibfnamefont{X.}~\bibnamefont{Hua}}, \bibnamefont{and}
  \bibinfo{author}{\bibfnamefont{H.}~\bibnamefont{Metcalf}},
  \bibinfo{journal}{J. Opt. Soc. Am. B} \textbf{\bibinfo{volume}{32}},
  \bibinfo{pages}{B75} (\bibinfo{year}{2015}{\natexlab{a}}).

\bibitem[{\citenamefont{Corder et~al.}(2015{\natexlab{b}})\citenamefont{Corder,
  Arnold, and Metcalf}}]{Corder}
\bibinfo{author}{\bibfnamefont{C.}~\bibnamefont{Corder}},
  \bibinfo{author}{\bibfnamefont{B.}~\bibnamefont{Arnold}}, \bibnamefont{and}
  \bibinfo{author}{\bibfnamefont{H.}~\bibnamefont{Metcalf}},
  \bibinfo{journal}{Phys. Rev. Lett.} \textbf{\bibinfo{volume}{114}},
  \bibinfo{pages}{043002} (\bibinfo{year}{2015}{\natexlab{b}}).

\bibitem[{\citenamefont{Shore}(1990)}]{Shore}
\bibinfo{author}{\bibfnamefont{B.}~\bibnamefont{Shore}},
  \emph{\bibinfo{title}{The Theory of Coherent Atomic Excitation}},
  vol.~\bibinfo{volume}{1} (\bibinfo{publisher}{Wiley, New York},
  \bibinfo{year}{1990}).

\bibitem[{\citenamefont{Chr\'etien}(2014)}]{Chretien}
\bibinfo{author}{\bibfnamefont{R.}~\bibnamefont{Chr\'etien}}, Master's thesis,
  \bibinfo{school}{Facult\'e des Sciences Appliqu\'ees, Universit\`e de
  Li\`ege}, \bibinfo{address}{Belgium} (\bibinfo{year}{2014}).

\bibitem[{\citenamefont{Steck}(2019{\natexlab{a}})}]{Steck:Na}
\bibinfo{author}{\bibfnamefont{D.~A.} \bibnamefont{Steck}},
  \emph{\bibinfo{title}{Sodium {D Line Data}}}
  (\bibinfo{year}{2019}{\natexlab{a}}),
  \bibinfo{note}{\url{https://steck.us/alkalidata/sodiumnumbers.pdf}}.

\bibitem[{\citenamefont{Steck}(2019{\natexlab{b}})}]{Steck:Cs}
\bibinfo{author}{\bibfnamefont{D.~A.} \bibnamefont{Steck}},
  \emph{\bibinfo{title}{Cesium {D Line Data}}}
  (\bibinfo{year}{2019}{\natexlab{b}}),
  \bibinfo{note}{\url{https://steck.us/alkalidata/cesiumnumbers.pdf}}.

\bibitem[{\citenamefont{Freund et~al.}(1975)\citenamefont{Freund, R\"omheld,
  and Oka}}]{Freund}
\bibinfo{author}{\bibfnamefont{S.~M.} \bibnamefont{Freund}},
  \bibinfo{author}{\bibfnamefont{M.}~\bibnamefont{R\"omheld}},
  \bibnamefont{and} \bibinfo{author}{\bibfnamefont{T.}~\bibnamefont{Oka}},
  \bibinfo{journal}{Phys. Rev. Lett.} \textbf{\bibinfo{volume}{35}},
  \bibinfo{pages}{1497} (\bibinfo{year}{1975}).

\bibitem[{\citenamefont{Kyr{\"o}la and Stenholm}(1977)}]{Kyrola77}
\bibinfo{author}{\bibfnamefont{E.}~\bibnamefont{Kyr{\"o}la}} \bibnamefont{and}
  \bibinfo{author}{\bibfnamefont{S.}~\bibnamefont{Stenholm}},
  \bibinfo{journal}{Optics Communications} \textbf{\bibinfo{volume}{22}},
  \bibinfo{pages}{123 } (\bibinfo{year}{1977}).

\end{thebibliography}

\end{document}